\documentclass[english,aps, prb,amsmath,amssymb,twocolumn,nofootinbib,superscriptaddress]{revtex4}
\usepackage[T1]{fontenc}
\usepackage[latin9]{inputenc}
\usepackage{color}
\usepackage{float}
\usepackage{amsbsy}
\usepackage{amstext}
\usepackage{graphicx}
\usepackage{esint}
\usepackage{epstopdf}
\usepackage{hyperref}

\usepackage{amsmath}

\makeatletter

\@ifundefined{textcolor}{}
{%
 \definecolor{BLACK}{gray}{0}
 \definecolor{WHITE}{gray}{1}
 \definecolor{RED}{rgb}{1,0,0}
 \definecolor{GREEN}{rgb}{0,1,0}
 \definecolor{BLUE}{rgb}{0,0,1}
 \definecolor{CYAN}{cmyk}{1,0,0,0}
 \definecolor{MAGENTA}{cmyk}{0,1,0,0}
 \definecolor{YELLOW}{cmyk}{0,0,1,0}
 }

\usepackage{dcolumn}
\usepackage{bm}
\usepackage{array}

\def\ket{\rangle}
\def\bra{\langle}

\newcommand{\PT}{\mathcal{PT}}

\newcommand{\beq}{\begin{equation}}
\newcommand{\eeq}{\end{equation}}
\newcommand{\beqa}{\begin{eqnarray}}
\newcommand{\eeqa}{\end{eqnarray}}

\makeatother

\usepackage{babel}

\begin{document}

\title{Anomalous-order exceptional point and non-Markovian Purcell effect at threshold in 1-D continuum systems}

\author{Savannah Garmon}

\email{sgarmon@p.s.osakafu-u.ac.jp}
\affiliation{Department of Physical Science, 
Osaka Prefecture University, 
Gakuen-cho 1-1, Sakai 599-8531, Japan}
\affiliation{Institute of Industrial Science, University of Tokyo, Kashiwa 277-8574, Japan}

\author{Gonzalo Ordonez}
\affiliation{Department of Physics and Astronomy, Butler University, Gallahue Hall, 4600 Sunset Avenue, Indianapolis, Indiana 46208, USA}

\author{Naomichi Hatano}
\affiliation{Institute of Industrial Science, University of Tokyo, Kashiwa 277-8574, Japan}

\begin{abstract}
For a system consisting of a quantum emitter coupled near threshold (band edge) to a one-dimensional continuum with a van Hove singularity in the density of states, we demonstrate general conditions such that a characteristic triple level convergence occurs directly on the threshold as the coupling $g$ is shut off.  For small $g$ values the eigenvalue and norm of each of these states can be expanded in a Puiseux expansion in terms of powers of $g^{2/3}$, which suggests the influence of a third-order exceptional point.  However, in the actual $g \rightarrow 0$ limit, only two discrete states in fact coalesce as the system can be reduced to a $2 \times 2$ Jordan block; the third state instead merges with the continuum.  Moreover, the decay width of the resonance state involved in this convergence is significantly enhanced compared to the usual Fermi golden rule, which is consistent with the Purcell effect.  However, non-Markovian dynamics due to the branch-point effect are also enhanced near the threshold.  Applying a perturbative analysis in terms of the Puiseux expansion that takes into account the threshold influence, we show that the combination of these effects results in quantum emitter decay of the unusual form $1 - C t^{3/2}$ on the key timescale during which most of the decay occurs. 
We then present two conditions that must be satisfied at the threshold for the anomalous exceptional point to occur: the density of states must contain an inverse square-root divergence and the potential must be non-singular.  We further show that when the energy of the quantum emitter is detuned from threshold, the anomalous exceptional point splits into three ordinary exceptional points, two of which appear in the complex-extended parameter space.  These results provide deeper insight into a well-known problem in spontaneous decay at a photonic band edge.
\end{abstract}

\maketitle




\global\long\def\braket#1#2{\left\langle #1\right. \left| #2 \right\rangle }

\global\long\def\bbrakket#1#2{\left\langle #1\right. \left\Vert #2\right\rangle }

\global\long\def\av#1{\left\langle #1 \right\rangle }

\global\long\def\tr{\text{Tr}}

\global\long\def\im{\text{Im}}

\global\long\def\re{\text{Re}}

\global\long\def\sign{\text{sgn}}

\global\long\def\abs#1{\left|#1\right|}

\section{introduction}
\label{sec:intro}

The physics of coalescing eigenstates at exceptional points \cite{Kato,BerryEP,HeissEP} have received attention in recent years in a variety of physical contexts \cite{Berry03,EPexpt1a,KGM08,LASM09,KoreaEP,Longhi10,Zheng10,PTCircuitExpt,Stone12,Peng14,GGH15,AluReview,LEP,BSUSM20}.  In part, this results from the fact that exceptional points can be associated with $\mathcal{PT}$-symmetry breaking in parity-time ($\mathcal{PT}$) symmetric systems \cite{BB98,RDM05,BenderRPP07} that have been studied as a potential non-Hermitian formulation of quantum mechanics \cite{BenderRPP07,BQZ01,MostaJMP02,BBJ02}, as well as in more applied contexts \cite{RDM05,KGM08,Zheng10,PTCircuitExpt,Peng14,GGH15,MECM08,Guo09,LonghiPRL09,LonghiLABS,CGS11,Christo12,Feng12,MostaPRA13,Hamid14,Konotop_review,Obuse16,PTQWexpt}.  
However, exceptional points also appear in traditional open quantum systems described by a Hermitian Hamiltonian that incorporates both discrete and continuous spectra \cite{Rotter_review,RotterBird}. In this context, exceptional points denote the appearance of resonances that are usually associated with exponential decay \cite{Gamow28,Siegert39,Sudarshan78,PPT91,HSNP08,Madrid12,OH17A,OH17B}, although at least in the immediate vicinity of the exceptional point 
the dynamics might still be non-exponential even when the resonance is present \cite{GO17}.

In general, the exceptional point (EP) represents a defective point in the parameter space of a given Hamiltonian where diagonalization is no longer possible as two or more eigenstates coalesce.  This situation is different than an ordinary degeneracy, at which the eigenvalues coincide while the corresponding eigenstates remain orthogonal to one another, and hence never coalesce.  In the usual picture, we refer to an exceptional point with $N$ coalescing eigenvalues as an EPN.  Then the simplest representation of the Hamiltonian at an EPN would contain an $N \times N$ Jordan block \cite{Kato,BS96,KGTP17}.  Meanwhile, in the vicinity of the EPN the eigenvalues (and other experimentally measurable quantities) can be expanded in terms of a {\it Puiseux series} of the form 
$E_j (\epsilon) = \bar{E} + \alpha_j (\epsilon - \bar{\epsilon})^{1/N} + \dots$ where
$\bar{\epsilon}$ is the exceptional point and $\bar{E}$ is the coalesced eigenvalue \cite{Kato,BerryEP,HeissEP,GGH15,Moiseyev}.  While most studies have focused on the simplest case involving the coalescence of two eigenstates \cite{Moiseyev1980,EPexpt1a,Hashimoto}, higher-order exceptional points involving three \cite{HeissEP3,CMW09,GraefeEP3,HW16,Heiss17,ChristoEP3,WiersigEP3} or more \cite{ZnojilEP4,JoglekarEP4,ZhongEPN} eigenstates have received more attention in the last few years \cite{GGKN08,UTKM13,LonghiEP3,HatanoEP3}.  
Proposed applications involving exceptional points include enhanced sensing \cite{Wiersig14,ChristoEP3,WiersigEP3,Stone19}
as well as modified spontaneous emission \cite{Lin16,Pick17,Dunham21}
and dynamical control \cite{GO17,LonghiEP3,Dietz07,Wiersig08,CM11,Reboiro19,Wiersig20,KBH21}.

In this paper, we analyze an exceptional point that generically occurs at the continuum threshold (band edge) when a quantum emitter is coupled to a one-dimensional (1-D) system under certain conditions.  As a result of the van Hove singularity in the 1-D density of states with characteristic divergence $1/\sqrt{E - E_\textrm{th}}$, in which $E_\textrm{th}$ is the continuum threshold, there occurs a nearby level triplet consisting of a resonance, an anti-resonance and a bound state.  In the vicinity of the threshold, their eigenstates and eigenvalues can be expanded in a Puiseux series in terms of $g^{2/3}$, in which $g$ is the coupling to the quantum emitter.  This seems to suggest an EP3 at which all three levels will coalesce in the limit $g \rightarrow 0$.  In fact, we find that 
the exceptional point is technically an EP2, despite behaving for physical purposes essentially like an EP3.
Further, the decay width $\Gamma$ of the resonance is enhanced near the threshold ($\Gamma \sim g^{4/3}$) in comparison to the usual case in which Fermi's golden rule could be applied ($\Gamma \sim g^2$) \cite{PTG05}.  This enhancement can be viewed as an implication of the Purcell effect, which predicts that the decay width of a quantum emitter is enhanced when it is tuned to the natural frequency of a cavity or waveguide \cite{Purcell,Kleppner,CQED}.

After establishing these properties of the exceptional point, we analyze in detail its influence on the survival probability of the excited quantum emitter state.  
During the critical timescale in which most of the decay occurs, a power-law evolution manifests itself in which the exponent is determined by two factors: one coming from the anomalous exceptional point and the other coming from the continuum threshold (branch-point effect) \cite{GO17,GPSS13}.  The dynamics on this timescale can be viewed as a non-Markovian correction to the previously-mentioned Purcell effect.
We emphasize that this effect should be rather universal; in particular, the same dynamics occurs in a well-known problem involving spontaneous emission near the edge of a photonic band gap  \cite{KKS94,John94,LNNB00}, but neither the power-law decay nor the exceptional point shaping those dynamics were previously noticed.

The models considered in this paper are written in terms of a microscopic Hamiltonian that includes a structured reservoir to describe the environmental influence on the quantum emitter, which means that there is a background continuum with a well-defined bandwidth and density of states  
\cite{John90,NBL99,DG03,LonghiPRA06,DBP08,GNHP09,BCP10,LZMM88,PTG05,KKS94,John94,LNNB00,CCCR16,SWGTC16,GTC2D1,GTC2D2,GTC3D}.
This is opposed to the coupled mode theory that has often been used to describe exceptional-point phenomena, in which one assumes that the essential physics can be described in terms of a few interacting resonance modes, while the microscopic details of the system are set aside (including the continuum).
Although the microscopic description applied here generally requires more effort to analyze the influence of the exceptional point, in some situations the properties of the discrete spectrum and those of the continuum cannot so neatly be disentangled \cite{GO17,KGTP17}.

In Sec. \ref{sec:inf} below we introduce a simple system consisting of a quantum dot and a tight-binding chain in order to explicitly demonstrate the properties of the threshold EP. 
In Sec. \ref{sec:inf.quad} we rewrite the model in terms of the quadratic eigenvalue problem, which enables us to describe the physical and mathematical properties of the exceptional point.  Then in Sec. \ref{sec:spec} we study the spectrum when the quantum emitter frequency is detuned from the threshold, observing several ordinary EP2s that occur in this vicinity.  As the detuning and the coupling are both shut off, these ordinary EP2s merge on the threshold as the previously discussed anomalous-order exceptional point appears.
We show the Jordan block structure and other technical properties of the anomalous-order exceptional point in Sec. \ref{sec:EP} [see also App. \ref{app:JB}].
Then in Sec. \ref{sec:time} we consider the influence of the EP on the time evolution of the excited quantum emitter state near the threshold.  We demonstrate that a $(1 - C t^{3/2})$-type power law decay occurs as a result of the influence of the anomalous exceptional point.  We discuss our results in Sec. \ref{sec:conc} and present our argument that two conditions should be satisfied for the anomalous-order EP to appear in a generic 1-D quantum system.  We also discuss a potential experiment in circuit quantum electrodynamics (QED).


\section{Model and formalism}\label{sec:inf}

The primary model that we consider in this paper is a quantum wire superlattice with an attached quantum dot \cite{TGP06}.  
The quantum wire is modeled by an infinite tight-binding array such that
our Hamiltonian takes the form
\begin{equation}
H= \epsilon_d d^\dag d
-J \sum_{j = -\infty}^{\infty}(c^\dag_j c_{j+1} + c^\dag_{j+1} c_{j})
-g \, (c_{0}^\dag d + d^\dag c_{0})
	,
\label{inf.ham}
\end{equation}
in which $c_j^\dagger$ is the creation operator at site $j$ of the array, while $J$ is the resonant coupling strength between array elements, which we set to unity as the unit of the energy.  Meanwhile $d^\dagger$ is the creation operator for the quantum dot excited state with excitation energy $\epsilon_d$.  The dot is coupled to the $0$ element of the chain $c_0^\dagger$ through the small coupling parameter $g$.
Note that because there is no particle-particle interaction in this Hamiltonian, we can study the dynamics within the single-particle subspace without approximation; further, the following analysis could equally well apply for an equivalent Bosonic realization of the model such as in waveguide QED in Ref. \cite{SZMG17}.


\subsection{Eigenvalues: Siegert boundary condition, effective Hamiltonian and dispersion equation}\label{sec:inf.Heff}

To analyze the spectrum of our model we next write our physical solution with the outgoing (or Siegert \cite{Siegert39}) boundary conditions
\begin{equation}
  \psi (x) \equiv \bra x | \psi \ket =
	\left\{ \begin{array}{ll}
		B e^{-ikx}					& \mbox{for $x \le -1$} ,     \\
		\psi_0					& \mbox{for $x = 0$} ,	\\
		\psi_\textrm{d}				& \mbox{for $x = \textrm{d}$} ,  	\\
		C e^{ikx}					& \mbox{for $x \ge 1$} .
	\end{array}
	\right.  
\label{psi.out}
\end{equation}
Writing the Schr\"odinger equation $\bra x | H | \psi \ket = E \bra x | \psi \ket$ far from the dot $|x| > 1$ yields the expression
\beq
  - \bra x + 1 | \psi \ket - \bra x - 1 | \psi \ket = E \bra x | \psi \ket
  	,
\label{sch.x}
\eeq
where we have set $J=1$.
Plugging into this 
Eq. (\ref{psi.out}) we find the continuous eigenvalue
\beq
  E_k = - 2 \cos k
  	,
\label{cont.disp}
\eeq
over the domain $k \in \left[ -\pi, \pi \right]$.
This defines the continuum 
in the range $E_k \in \left[-2, 2 \right]$ with the associated density of states given by
\beq
  \rho (E) = \frac{\partial k}{\partial E}
  		= \frac{1}{\sqrt{4-E^2}}
	.
\label{inf.DOS}
\eeq
The divergences occurring at the band edges (continuum threshold) $E= \pm 2$ are the van Hove singularities characteristic of 1-D systems \cite{vanHove,Mahan,Economou}.

Next we aim to obtain the discrete spectrum associated with the quantum dot by solving the Schr\"odinger equation in its vicinity.  To do so we first introduce the projection operator 
$P = | 0 \ket \bra 0 | + | d \ket \bra d |$ for the central region consisting of the dot $| d \ket$ and the chain site $| 0 \ket$ to which it couples.  We further introduce the operator $Q = 1 - P$ that represents the system environment.
We then project out the $Q$ sector according to the Feshbach method  \cite{Feshbach1,Feshbach2} to write the Schr\"odinger equation in the $P$ sector as
\beq
   H_\textrm{eff} (E_j) P | \psi_j \ket = E_j \left( P | \psi_j \ket \right)
   	,
\label{H_eff.eqn}
\eeq
where we have applied Eq. (\ref{psi.out}).
Here, the effective Hamiltonian $H_\textrm{eff} (E)$ is given by
\beqa
  H_\textrm{eff} (E)
    & 	= & PHP + PHQ \frac{1}{E - E_k} QHP
			\nonumber		\\
   &  	= & \left( \begin{array}{ccc}
		- 2 e^{i k}	& - g    	\\
		- g		& \epsilon_d
	\end{array}
	\right)
	;
\label{inf.H_eff}
\eeqa
see Appendices of Refs. \cite{SHO11,HO14}.
We emphasize that $H_\textrm{eff}$ depends on $E$ through Eq. (\ref{cont.disp}).  Now, taking the discriminant of Eq. (\ref{H_eff.eqn}), we obtain the dispersion equation for the discrete eigenvalues in the form
\beq
  E - \epsilon_d
  	= \Sigma (E) = \left\{ \begin{array}{cc} 
					- \frac{g^2}{\sqrt{E^2 - 4}} 	&   \textrm{for} 	\ E < -2 ,	\\
					- \frac{g^2}{i \sqrt{4 - E^2}} 	&   \textrm{for}	\ | \re \ E | < -2, \im \ E = 0^+ ,  \\
					\frac{g^2}{\sqrt{E^2 - 4}} 	&   \textrm{for}	\ E > 2 ,
		 \end{array} \right.
\label{inf.self}
\eeq
where $\Sigma (E)$ has been written specifically in the first Riemann sheet and can be analytically continued to the second sheet by the usual methods.
After squaring to remove the root, we obtain an equivalent quartic polynomial condition for the eigenvalues
as $p(E_j) = 0$, in which 
\beq
  p(E) = \left(E - \epsilon_d \right)^2 \left(E^2 - 4 \right) - g^4
  	.
\label{inf.poly}
\eeq

\begin{figure*}
\hspace*{0.05\textwidth}
 \includegraphics[width=0.4\textwidth]{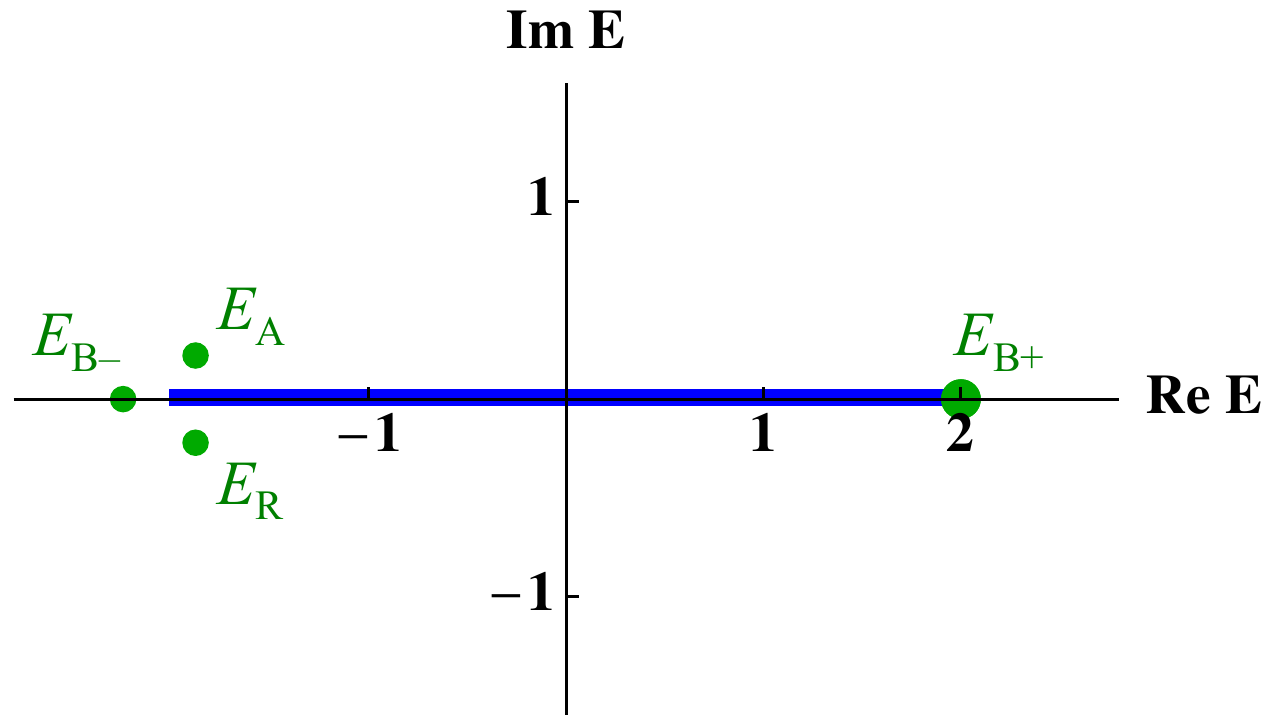}
\hfill
 \includegraphics[width=0.4\textwidth]{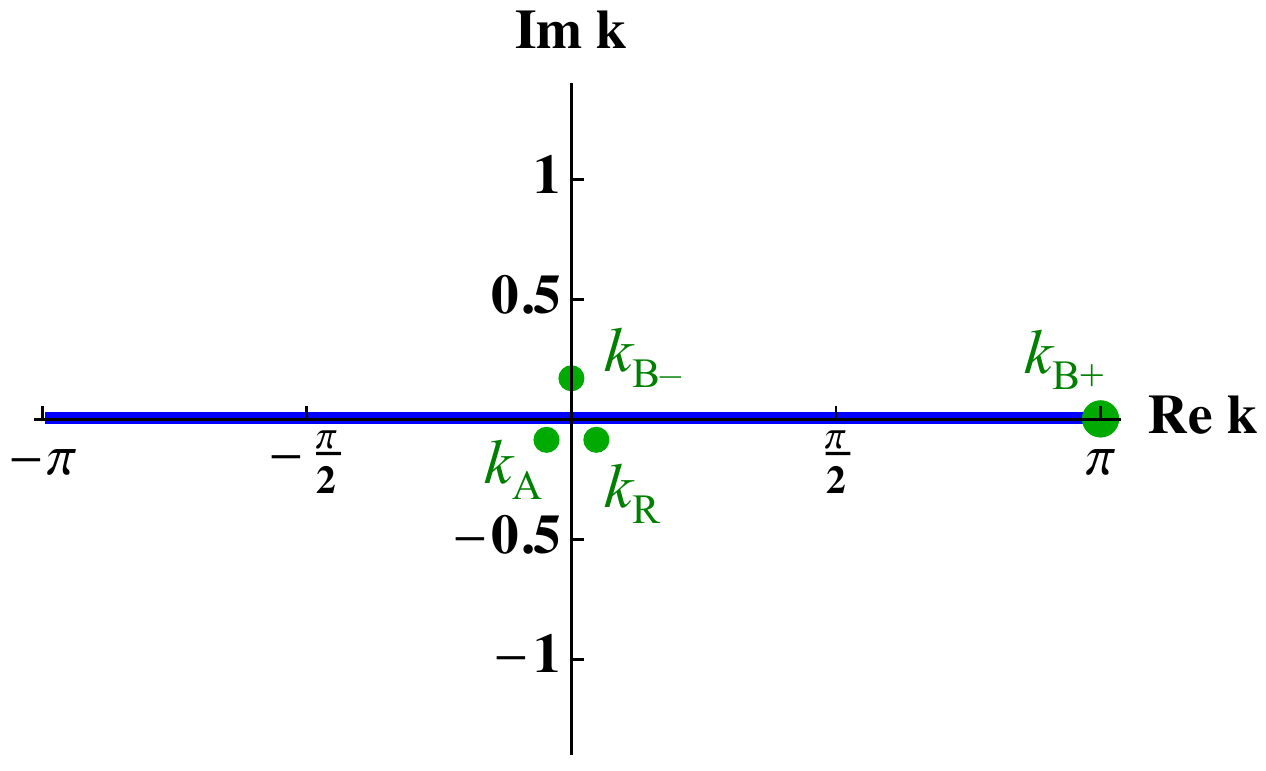}
 \hspace*{0.05\textwidth}
\\
\hspace*{0.01\textwidth}(a)\hspace*{0.47\textwidth}(b)\hspace*{0.4\textwidth}
\\
\caption{The four discrete eigenvalues in (a) the complex $E$ plane, and (b) the complex $k$ plane, for the case  $\epsilon_d = - 2$ and $g = 0.5$.  The upper bound state $E_{\textrm{B}+}$ eigenvalue is indicated with a larger marker for visualization purposes only.  (Energy is measured in units of $J=1$ and the wave number is measured in units $a=1$ throughout the paper.)
}
\label{fig:E.k.plane} 
\end{figure*}

Let us point out an immediate consequence of Eq. (\ref{inf.poly}).  Notice that if we take $\epsilon_d = -2$, the polynomial becomes $p(E) = (E + 2)^3 (E-2) - g^4 = 0$.  If $g < 1$ is small, this results in three eigenvalues clustered near the lower band edge at $E = -2$. 
We present these eigenvalues as they appear in the complex energy plane for the case $g = 0.5, \epsilon_d = -2$ in Fig. \ref{fig:E.k.plane} (a).  We further show the associated value of the complex wave vector $k_j$, obtained from the dispersion $E_j = - 2 \cos k_j$, for each eigenvalue in Fig. \ref{fig:E.k.plane} (b).  Notice that one of the three clustered eigenvalues ($E_{\textrm{B}-}$) has $\im \ k_{\textrm{B}-} > 0$, for which the wave function form Eq. (\ref{psi.out}) yields a normalizable solution, implying that $E_{\textrm{B}-}$ represents a bound state appearing in the first Riemann sheet of the complex energy plane.  The other two among the clustered eigenvalues are a resonance $E_\textrm{R}$ and an anti-resonance $E_\textrm{A}$ with complex conjugate eigenvalues and non-normalizable wave functions (these reside in the second sheet).
Finally, the fourth eigenvalue  $E_{\textrm{B}+}$ is also a bound state but instead appears near the upper band edge.  However, this fourth solution plays little role in the physics in this situation, and hence we can mostly ignore it in what follows.

\begin{figure}
\includegraphics[width=0.9\columnwidth]{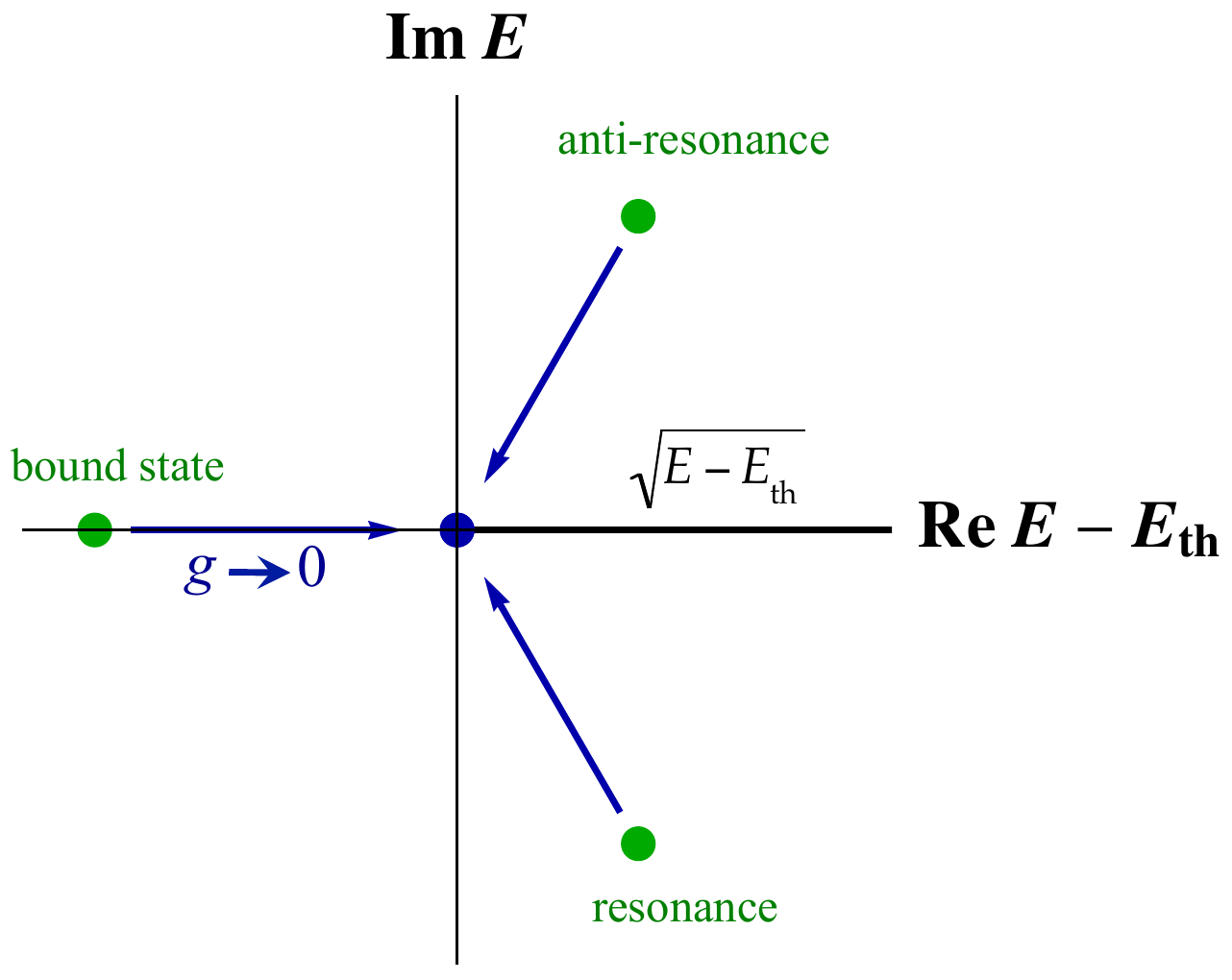} 
\caption{Diagram demonstrating the convergence of the three eigenvalues on the band edge (continuum threshold) $E = E_\textrm{th}$ as $g \rightarrow 0$ for a generic model.  This corresponds to the case $\epsilon_d = -2$ and $E_\textrm{th} = -2$ (with $J=1$) for the model in the main text.
}
\label{fig:plane.EP}
\end{figure}

From the form of the polynomial $p(E) = (E + 2)^3 (E-2) - g^4 = 0$ at $\epsilon_d = -2$, we can also easily see that in the limit $g \rightarrow 0$, the three clustered eigenvalues must converge on the lower band edge.  This behavior is illustrated diagrammatically in Fig. \ref{fig:plane.EP}.  Further, we can obtain an expansion for the eigenvalues in the vicinity of the band edge (threshold) for small, nonzero $g$ values by 
substituting an ansatz of the form 
$E_{\textrm{B}-} = -2 + \chi_\alpha g^{\alpha} + \chi_\beta g^{\beta} + \dots$ into the dispersion Eq. (\ref{inf.poly}) and solving for the coefficients $\chi_s$ and the exponents.  Doing so, we find an expansion for the bound-state eigenvalue
\beq
  E_{\textrm{B}-}
  	= -2 - \frac{g^{4/3}}{2^{2/3}} + \frac{g^{8/3}}{24 \times 2^{1/3}} + \mathcal{O}(g^{4})
	.
\label{E.B}
\eeq
as well as the resonance and anti-resonance eigenvalues
\beq
  E_\textrm{R,A}
  	= -2 + \frac{e^{\pm \pi i /3} g^{4/3}}{2^{2/3}} 
			+ \frac{e^{\mp \pi i /3} g^{8/3}}{24 \times 2^{1/3}} + \mathcal{O}(g^{4})
	.
\label{E.RA}
\eeq
The resonance width $\Gamma = 2 \im \ E_\textrm{R}$ has the predicted form $\Gamma \sim g^{4/3}$, which is the Markovian aspect of the Purcell effect discussed in Sec . \ref{sec:intro}.

\subsection{Eigenstates: Mapping to the quadratic eigenvalue problem}\label{sec:inf.quad}

While the preceding formalism has yielded the energy eigenvalue spectrum for the model in Eq. (\ref{inf.ham}), notice that it is not quite adequate to fully describe the behavior of the three apparently converging eigenstates in the $g \rightarrow 0$ limit, because in Eq. (\ref{inf.H_eff}) the effective Hamiltonian is written as a function of its own eigenvalue, which means that any two differing eigenstates 
technically belong to the solution space of two different copies of $H_\textrm{eff}$.  In other words, Eq. (\ref{H_eff.eqn}) represents a nonlinear eigenvalue problem (See Sec. II. C. of Ref. \cite{GO17} for a similar discussion).  To address this problem, notice that we could equivalently write the effective Hamiltonian Eq. (\ref{inf.H_eff}) in terms of the variable $\lambda = e^{i k}$.  This is a convenient choice because we can use $E = - 2 \cos k$ to write the energy variable appearing on the right-hand side of Eq. (\ref{H_eff.eqn}) as
\beq
  E = - \lambda - \frac{1}{\lambda}
  	.
\label{cont.disp.lambda}
\eeq
Taken together, this allows us to rewrite Eq. (\ref{H_eff.eqn}) as an equivalent quadratic eigenvalue problem in $\lambda$ in the form of the coupled equations
\beq
  \left( 1 - \lambda^2 \right) \bra 0 | \psi \ket
  		- g \lambda  \bra d | \psi \ket = 0
\label{inf.quad.1}
\eeq
and
\beq
  - g \lambda \bra 0 | \psi \ket
  		 +  \left( \lambda^2 + \epsilon_d \lambda + 1 \right) \bra d | \psi \ket = 0
	.
\label{inf.quad.2}
\eeq
In general, the quadratic eigenvalue problem can be solved after mapping to a generalized linear eigenvalue problem \cite{SIAM,quad_book}.  In the present case, we do so by rewriting the paired equations (\ref{inf.quad.1}) and  (\ref{inf.quad.2}) in the form
\beq
  \left( A - \lambda B \right) | \Psi \ket = 0
\label{GEP.eqn}
\eeq
in which
\beq
  A = \left[ \begin{array}{cccc}
  		0 &	0 &	1 &	0	\\
		0 &	0 &	0 &	1	\\
		1  &	0 &	0 &	-g	\\
		0 &	1 &	-g &	\epsilon_d
	\end{array}
	\right]
	,
	\ \ \ \ \ \ \ \ \ \ \ \ \ \ \ \ \ \ \ 
  B = \left[ \begin{array}{cccc}
  		1 &	0 &	0 &	0	\\
		0 &	1 &	0 &	0	\\
		0  &	0 &	1 &	0	\\
		0 &	0 &	0 &	-1
	\end{array}
	\right]
	,
\label{inf.A.B}
\eeq
and the eigenvector $ | \Psi \ket$ takes the form
\beq
  | \Psi \ket
  	= \left[ \begin{array}{c}
		\bra 0 | \psi \ket			\\
		\bra d | \psi \ket			\\
		\lambda \bra 0 | \psi \ket	\\
		\lambda \bra d | \psi \ket
	\end{array}
	\right]
	;
\label{inf.Psi}
\eeq
see Ref. \cite{HO14}.
The four discrete eigenvalues $\lambda_j$ are obtained in the present context from the discriminant of Eq. (\ref{GEP.eqn}), which yields the quartic $f(\lambda_j) = 0$ with
\beq
  f(\lambda) 
  	=  - \lambda^4 - \epsilon_d \lambda^3 - g^2 \lambda^2 + \epsilon_d \lambda + 1
	.
\label{inf.poly.lambda}
\eeq
This is equivalent to the quartic for the energy eigenvalue $p(E_j) = 0$ in Eq. (\ref{inf.poly}).
The norm of the corresponding eigenstates is fixed by the normalization condition 
$\bra \tilde{\Psi}_j | B | \Psi_j \ket = 1$, which gives
\beq
  \left( 1 + \lambda_j^2 \right) \bra 0 | \psi_j \ket^2 + \left( 1 - \lambda_j^2 \right) \bra d | \psi_j \ket^2
  	= 1
	.
\label{inf.norm}
\eeq
Combining this with Eq. (\ref{inf.quad.1}) we obtain the conditions
\beq
  \bra 0 | \psi_j \ket
  	= \frac{g \lambda_j}
		{\sqrt{g^2 \lambda_j^2 \left( 1 + \lambda_j^2 \right) + \left( 1 - \lambda_j^2 \right)^3 }}
\label{inf.norm.0}
\eeq
and
\beq
  \bra d | \psi_j \ket
  	= \frac{ 1 - \lambda_j^2}
		{\sqrt{g^2 \lambda_j^2 \left( 1 + \lambda_j^2 \right) + \left( 1 - \lambda_j^2 \right)^3 }}
	.
\label{inf.norm.d}
\eeq

With the present formalism in hand, the four eigenstate solutions $|\Psi_j \ket$ from Eq. (\ref{GEP.eqn}) are now in one-to-one correspondence with the four discrete eigenvalues from $f(\lambda_j) = 0$.
This will enable us to study precisely the behavior of the eigenvectors as we approach the triple degeneracy at the threshold in Sec. \ref{sec:JB} and App. \ref{app:JB} as well as the influence of the degeneracy on the survival probability of the occupied dot state in Sec. \ref{sec:time}.  Before turning to these issues, however, we first examine the eigenvalue spectrum in the vicinity of the threshold in greater detail in the next section.


\section{Eigenvalue spectrum in the vicinity of the threshold} \label{sec:spec}


\begin{figure*}
\hspace*{0.05\textwidth}
 \includegraphics[width=0.4\textwidth]{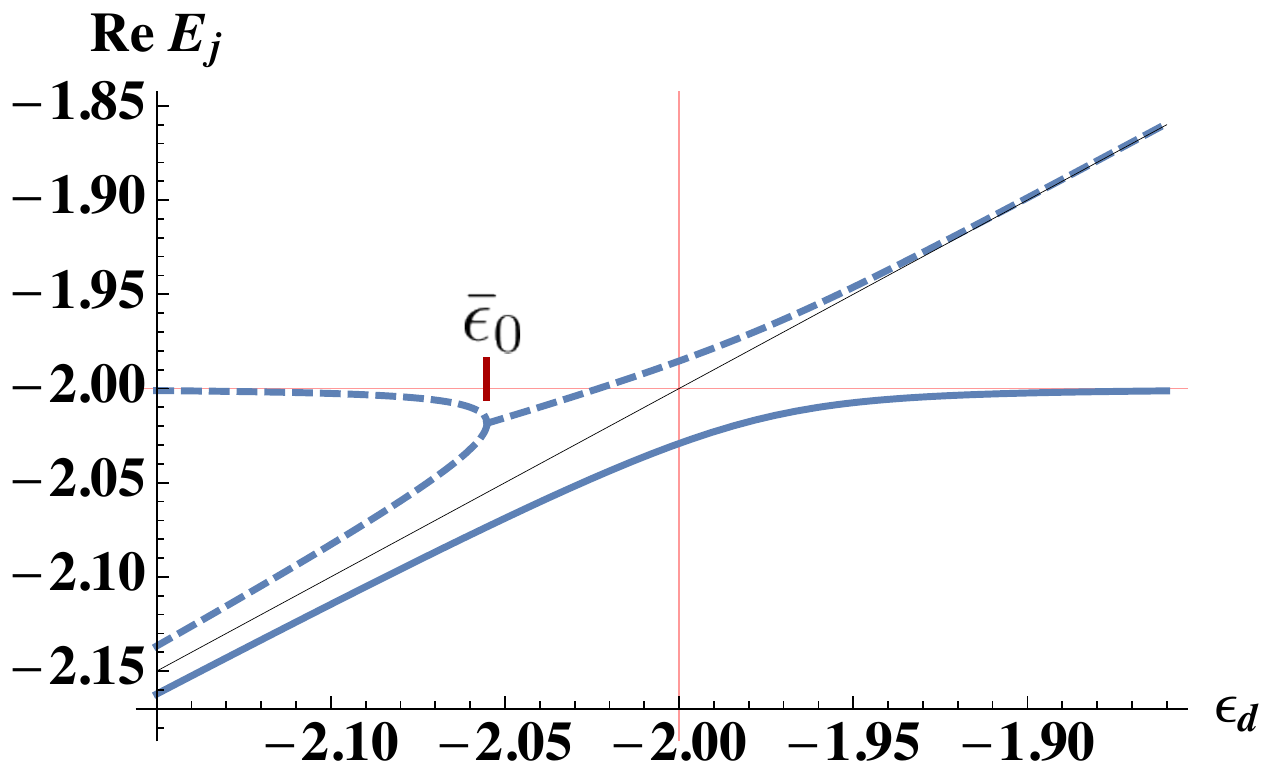}
\hfill
 \includegraphics[width=0.4\textwidth]{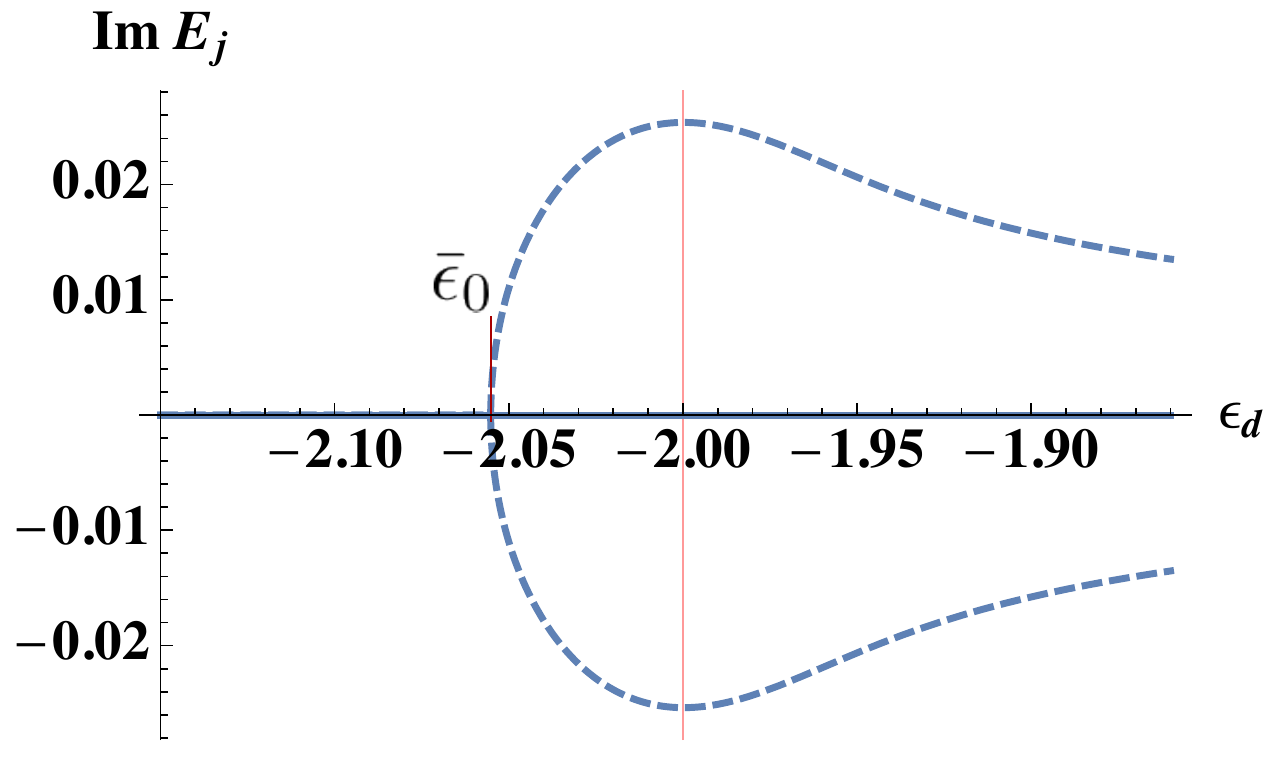}
 \hspace*{0.05\textwidth}
\\
\vspace*{\baselineskip}
\hspace*{0.01\textwidth}(a)\hspace*{0.47\textwidth}(b)\hspace*{0.4\textwidth}
\\
\caption{(a) Real part and (b) imaginary part of the three eigenvalues appearing near the lower band edge for $g=0.1$ and $\epsilon_d$ in the vicinity $\epsilon_d \sim -2$.  The solid curve indicates the bound state $E_{\textrm{B}-}$ in the first Riemann sheet, while the dashed curves indicates the two solutions in the second sheet.
}
\label{fig:inf.spec} 
\end{figure*}

In the previous section, we saw a snapshot of the spectrum for the case that $\epsilon_d$ coincided with the lower band edge (threshold) $E = -2$. In this section, we relax this condition and allow $\epsilon_d$ to vary in the immediate vicinity of the band edge in order to illustrate the nearby presence of three ordinary exceptional points.  We will locate these EPs by applying the method from Ref. \cite{GRHS12}.

\begin{figure}
\includegraphics[width=0.92\columnwidth]{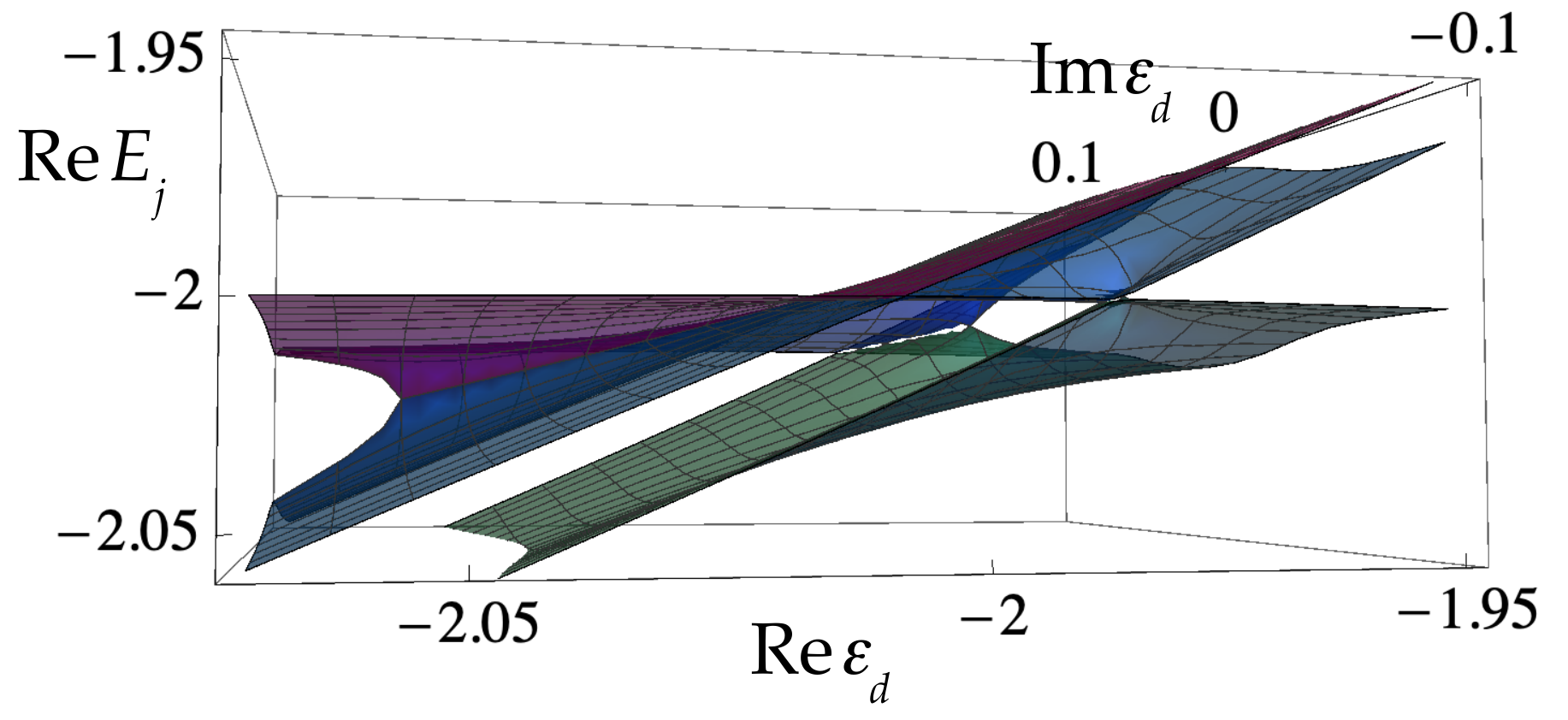} 
\\
\hspace*{0.04\columnwidth}(a)\hspace*{0.96\columnwidth}
\\
 \includegraphics[width=0.92\columnwidth]{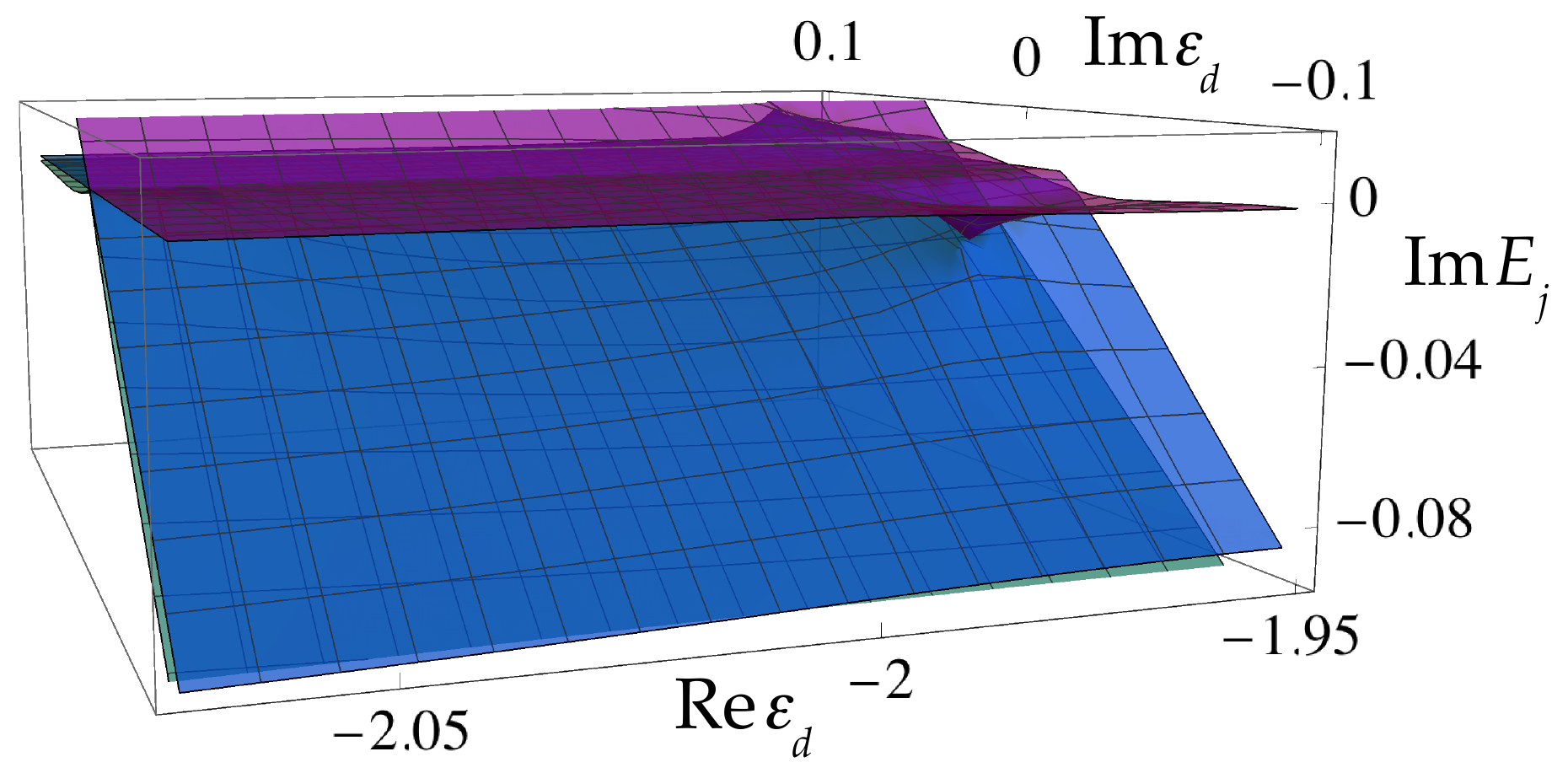}
 \\
\hspace*{0.04\columnwidth}(b)\hspace*{0.96\columnwidth}
\\
 \includegraphics[width=0.92\columnwidth]{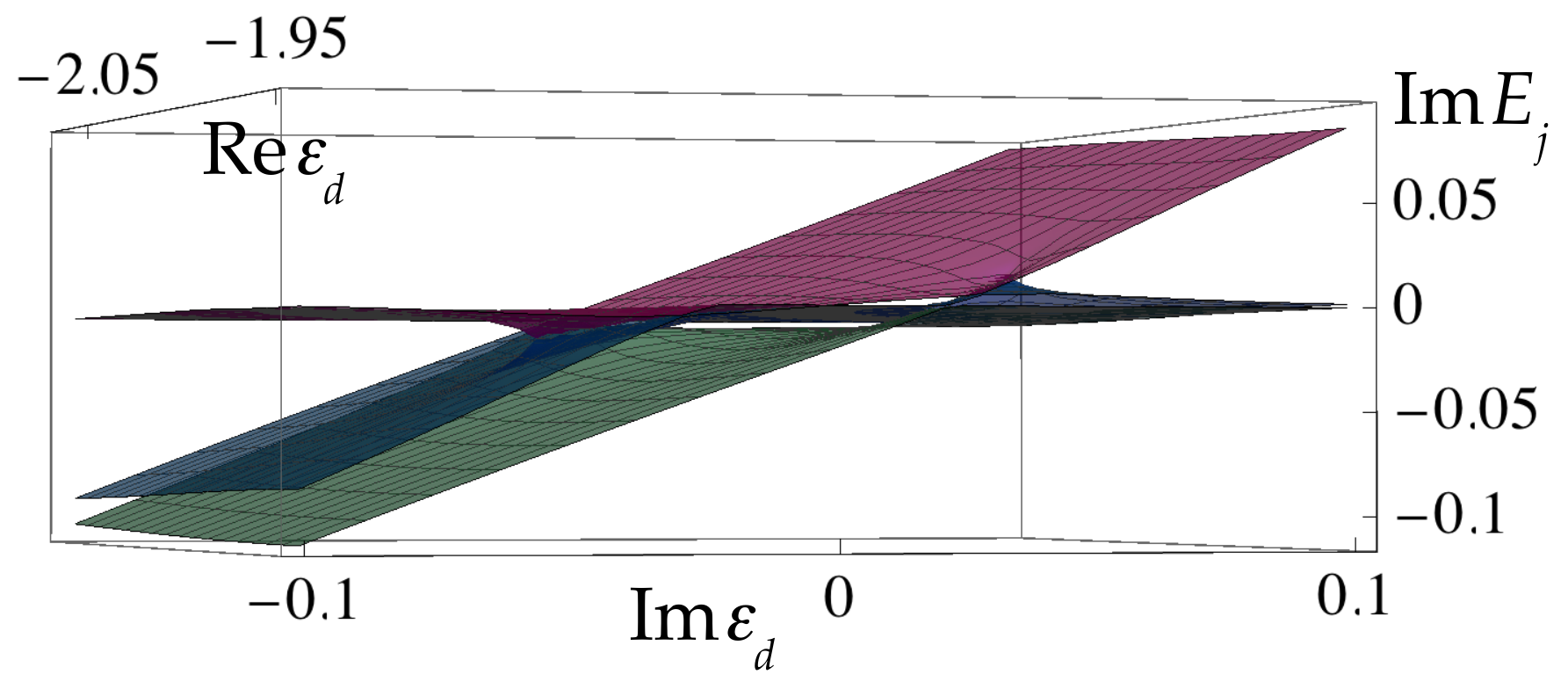}
  \\
\hspace*{0.04\columnwidth}(c)\hspace*{0.96\columnwidth}
\caption{Extension of the (a) real parts and (b, c) imaginary parts of the three eigenvalues appearing near the lower band edge in the complex-extended  $\epsilon_d$ plane.  We emphasize that (b) and (c) show the imaginary parts of the same eigenvalues from two different viewpoints. (Energy is measured in units of $J=1$ throughout the paper.)
}
\label{fig:inf.comp.EP2s}
\end{figure}

In Fig. \ref{fig:inf.spec}(a) we plot the real part of the three eigenvalues that appear near the lower band edge
for the value $g = 0.1$ while $\epsilon_d$ varies over $\epsilon_d \in \left[ -2.15, -1.85 \right]$.
The solid curve here indicates the bound state $E_-$.
For values of $\epsilon_d < -2$, this bound state appears shifted slightly below the unperturbed energy $\epsilon_d$ itself.  However, for $\epsilon_d > -2$ this eigenvalue breaks off and gets ``stuck,'' always appearing below the lower band edge.
Meanwhile, the two other eigenvalues are shown by a dashed curve.
When $\epsilon_d$ appears well below the continuum at $\epsilon_d < \bar{\epsilon}_0 \approx -2.05518 \dots$ these solutions appear as two {\it virtual bound states}, which are non-normalizable states with real eigenvalue in the second Riemann sheet 
\cite{Moiseyev,HO14,Nussenzveig,BCP10,GPSS13,GNOS19}
However, as $\epsilon_d$ draws closer to the continuum at $\epsilon_d = \bar{\epsilon}_0$ the virtual bound states coalesce before forming a resonance, anti-resonance pair with complex conjugate eigenvalues.  This shows that $\epsilon_d = \bar{\epsilon}_0$ is an (ordinary) EP2.  

We note that the above picture for the real part of the eigenvalues [see Fig. \ref{fig:inf.spec}(a)]  is reminiscent of the traditional avoided crossing picture in quantum systems; however, the resonance and anti-resonance actually collide while the bound state experiences an avoided crossing with the other two solutions.  We will momentarily obtain deeper insight into this picture.

The corresponding imaginary parts of these eigenvalues are then shown in Fig. \ref{fig:inf.spec}(b), in which we immediately see that the resonance decay width $- \im E_R \sim g^{4/3}$ for $\epsilon_d \gtrsim \bar{\epsilon}_0$ is enhanced near the threshold $\epsilon_d \approx -2$ compared to the case as $\epsilon_d$ becomes embedded deeper inside the continuum where the Fermi golden rule would apply.
This indicates that the relaxation process should be enhanced near the continuum threshold, as we would expect from the Purcell effect.  However, to get a full picture for the decay dynamics we will also have to take into account the branch-point effect; this is worked out later in Sec. \ref{sec:time}.

Now we aim to obtain precise information about the location of the EP2 at which the resonance and anti-resonance solutions appear.  In the vicinity of any exceptional point we can expand the associated eigenvalues in the characteristic Puiseux expansion \cite{Kato,Moiseyev1980,GRHS12}, which in the present case we expect takes the form $E_j = \bar{E} + \alpha_j (\epsilon_d - \bar{\epsilon}_0)^{1/2} + \dots$ with $\bar{E}$ being the coalesced eigenvalue and $\bar{\epsilon}_0$ the location of the exceptional point itself.  Notice from this expression that the derivative of the eigenvalue with respect to the parameter $\epsilon_d$ blows up at the exceptional point as
\beq
  \left. \frac{\partial E_j}{\partial \epsilon_d}  \right|_{\epsilon_d = \bar{\epsilon}_0} \rightarrow \infty
  	.
\label{EP.derivative}
\eeq
We make use of this fact to extract information about the EP in the following quick derivation.  First we take a full derivative of Eq. (\ref{inf.self}) and rearrange to obtain 
\beq
  1 - \frac{1}{\partial E / \partial \epsilon_d} = - \frac{g^2 E}{\left( E^2 - 4 \right)^{3/2}}
  	.
\eeq
Then letting $\epsilon_d \rightarrow \bar{\epsilon}_0$ (so that $E \rightarrow \bar{E}$) 
we obtain a condition on the coalesced eigenvalue as
\beq
  \left( \bar{E}^2 - 4 \right)^3 = g^4 \bar{E}^2
  	,
\label{inf.EP2.cond}
\eeq
This double-cubic polynomial yields three positive solutions (associated with the upper band edge) and three negative solutions (associated with the lower band edge).  Maintaining our focus on the lower band edge, 
we obtain the exact form of the three negative solutions as
\begin{widetext}
\begin{equation}
  \bar{E}_n
  	= - \sqrt{ 4 + e^{\frac{2 \pi i n }{ 3 }} g^{4/3} \left( 2 + \frac{1}{3} \sqrt{36 - \frac{g^4}{3}} \right)^{1/3}
			+ e^{\frac{4 \pi i n }{ 3 }} g^{4/3} \left( 2 - \frac{1}{3} \sqrt{36 - \frac{g^4}{3}} \right)^{1/3}
			}
	,
\label{inf.EP2.E}
\end{equation}
\end{widetext}
for $n = -1, 0, 1$.  The solution $\bar{E}_0$ corresponds to the previously-discussed exceptional point $\epsilon_d = \bar{\epsilon}_0$ at which the resonance and anti-resonance pair appear [see Fig. \ref{fig:inf.spec}(a)].  Meanwhile  $\bar{E}_{1}$ and $\bar{E}_{-1}$ are two complex-valued solutions of Eq. (\ref{inf.EP2.cond}) that can only be seen by extending the problem into the complex $\epsilon_d$ parameter space, as shown in Fig. \ref{fig:inf.comp.EP2s}.  Notice that whereas it was the resonance and anti-resonance that coalesced at the real-valued EP2 $\bar{\epsilon}_0$ corresponding to $\bar{E}_0$, it is instead the complex-extension of the bound state that coalesces with one of the other two eigenvalues at the complex-valued EP2s corresponding to $\bar{E}_1$ and $\bar{E}_{-1}$, while the third eigenvalue experiences an avoided crossing.  
(We note this configuration of three EP2s is also studied from a pure mathematics perspective in a recent work, Ref. \cite{FP21}.)

Also notice from Eq. (\ref{inf.EP2.E}) that for $g \rightarrow 0$ the three ordinary EP2s converge to the band edge at $E = -2$.  This suggests that in the $g \rightarrow 0$ limit the three ordinary EP2s collectively give rise to the anomalous threshold EP that was previously introduced at the end of Sec. \ref{sec:inf.Heff}, and which is the main topic of this work.  From this point forward, we focus our attention on the anomalous EP itself.



\section{Properties of the anomalous threshold exceptional point} \label{sec:EP}

For the remainder of the paper, we focus directly on the properties of the anomalous exceptional point at the lower band edge.  Hence, we now assume the parameter condition $\epsilon_d = -2$ is always satisfied, which corresponds to the vertical red line in Fig. \ref{fig:inf.spec}.


\subsection{Small $g$ behavior of the eigenstates near threshold} \label{sec:EP.g}

Previously, we obtained Puiseux expansions for the three converging energy eigenvalues in Eqs. (\ref{E.B}) and (\ref{E.RA}), which describe their behavior for small values of $g$ at $\epsilon_d = -2$.  Next we obtain similar expansions for the $\lambda$ eigenvalues introduced in Sec. \ref{sec:inf.quad} as well as the corresponding norm of the converging eigenstates.  These expressions will prove useful in analyzing the time evolution near the threshold later in Sec. \ref{sec:time}, while also providing insight into the mathematical properties of the anomalous exceptional point.

First, similar to the process that we used to obtain Eqs. (\ref{E.B}) and (\ref{E.RA}), we can obtain Puiseux expansions for the $\lambda$ eigenvalues for the bound state
\beq
  \lambda_{\textrm{B}-}
  	= 1 - \frac{g^{2/3}}{2^{1/3}} + \frac{g^{4/3}}{2^{5/3}} - \frac{g^2}{24} - \frac{g^{8/3}}{48 \times 2^{1/3}}
	+ \mathcal{O}(g^{10/3})
\label{lambda.B}
\eeq
and resonance and anti-resonance solutions
\beqa
  \lambda_\textrm{R,A}
  &	= & 1 + \frac{e^{\pm \pi i /3} g^{2/3}}{2^{1/3}} - \frac{e^{\mp \pi i /3} g^{4/3}}{2^{5/3}} 
						\nonumber	\\
  & &		- \frac{g^2}{24} + \frac{e^{\pm \pi i /3} g^{8/3}}{48 \times 2^{1/3}} +  \mathcal{O}(g^{10/3})
\label{lambda.RA}
\eeqa
after applying a generic form of these expansions in the $\lambda$ polynomial Eq. (\ref{inf.poly.lambda}). 

As our next step, we can apply the above expressions in Eq. (\ref{inf.norm.d}) to obtain expansions for the $\bra d |$ component of the norm in the vicinity of the anomalous EP.  For the bound state we obtain
\beq
  \bra d | \psi_{\textrm{B}-} \ket^2
  	= \frac{2^{1/3}}{3 g^{2/3}} + \frac{1}{3} + \frac{g^{2/3}}{9 \times 2^{1/3}} + \mathcal{O}(g^{4/3})
\label{norm.B.d}
\eeq
and for the resonance, anti-resonance pair we find
\beq
  \bra d | \psi_\textrm{R,A} \ket^2
  	= - \frac{e^{\mp \pi i /3} 2^{1/3}}{3 g^{2/3}} + \frac{1}{3} 
			- \frac{e^{\pm \pi i /3} g^{2/3}}{9 \times 2^{1/3}} + \mathcal{O}(g^{4/3})
	.
\label{norm.RA.d}
\eeq
Notice in Eqs. (\ref{norm.B.d}) and (\ref{norm.RA.d}) that the component of the norm $\bra d | \psi_{\textrm{B}-,\textrm{R,A}} \ket$ diverges for all three states in the limit $g \rightarrow 0$, which conforms with the expected behavior of an exceptional point \cite{Moiseyev,GO17,KGTP17}.  In particular, it again seems to suggest that all three of these states are converging to the dot state $| d \ket$ in this limit.  However, as we will show next, connecting the small $g$ behavior of the system with the actual limit $g \rightarrow 0$ is a more subtle issue than what this picture suggests.


\subsection{Jordan block structure in the $g \rightarrow 0$ limit} \label{sec:JB}

We now obtain the explicit form of the eigenvalue problem (\ref{GEP.eqn}) in the limit $g \rightarrow 0$.  To simplify our discussion somewhat, we first rewrite the generic form of the eigenvalue equation slightly as 
\beq
  \left( B^{-1} A - \lambda \right) | \Psi \ket = 0
  	,
\label{GEP.eqn.2}
\eeq
which takes the form of an ordinary linear eigenvalue problem of the non-Hermitian matrix $B^{-1} A$.
In the limit of interest $g \rightarrow 0$ with $\epsilon_d = -2$, the non-Hermitian matrix $B^{-1} A$ takes the form
\beq
  B^{-1} A
  	= \left( \begin{array}{cccc}
		0	& 0	& 1	& 0		\\
		0	& 0	& 0	& 1		\\		
		1	& 0	& 0	& 0		\\
		0	& -1	& 0	& 2					
		\end{array} \right)
\label{BA}
\eeq
It is then straightforward to show that this matrix can be transformed to the simplest form
\beq
  R^{-1} \left( B^{-1} A \right) R
  	= \left( \begin{array}{cccc}
		-1	& 0	& 0	& 0		\\
		0	& 1	& 1	& 0		\\		
		0	& 0	& 1	& 0		\\
		0	& 0	& 0	& 1					
		\end{array} \right)
	,
\label{JB}
\eeq
which contains a $2 \times 2$ Jordan block, plus an additional degeneracy (the form of $R$ is reported in App. \ref{app:JB}).  This unambiguously demonstrates that the exceptional point at the threshold is in fact an EP2, despite essentially behaving like an EP3 in the vicinity of the coalescence.  

More precisely connecting the EP3-like behavior for small $g$ with the actual EP2 in the $g \rightarrow 0$ limit can be accomplished by realizing that the three states $| \Psi_{\textrm{B}-} \ket$, $| \Psi_\textrm{R} \ket$, and $| \Psi_\textrm{A} \ket$ no longer really exist individually in this limit.  Instead, certain linear combinations can be taken to appropriately connect the spectrum in the two cases.  However, since the details are somewhat tedious and not strictly necessary for the proceeding analysis, we leave these details to App. \ref{app:JB}.


\section{Non-Markovian Purcell effect and influence of the EP} \label{sec:time}

Having established the spectral and mathematical properties of the anomalous exceptional point, we now analyze its influence on the relaxation process of the excited quantum emitter state.  A number of works have established that exceptional points can influence spontaneous emission in a variety of circumstances \cite{Lin16,Pick17,Dunham21,GO17,LonghiEP3,Dietz07,Wiersig08,CM11,Reboiro19,Wiersig20,KBH21}.  Below, we demonstrate that the combined influence of the anomalous exceptional point and the threshold itself determine the decay dynamics of the dot during the most consequential timescale.

\begin{figure*}
\hspace*{0.05\textwidth}
 \includegraphics[width=0.3\textwidth]{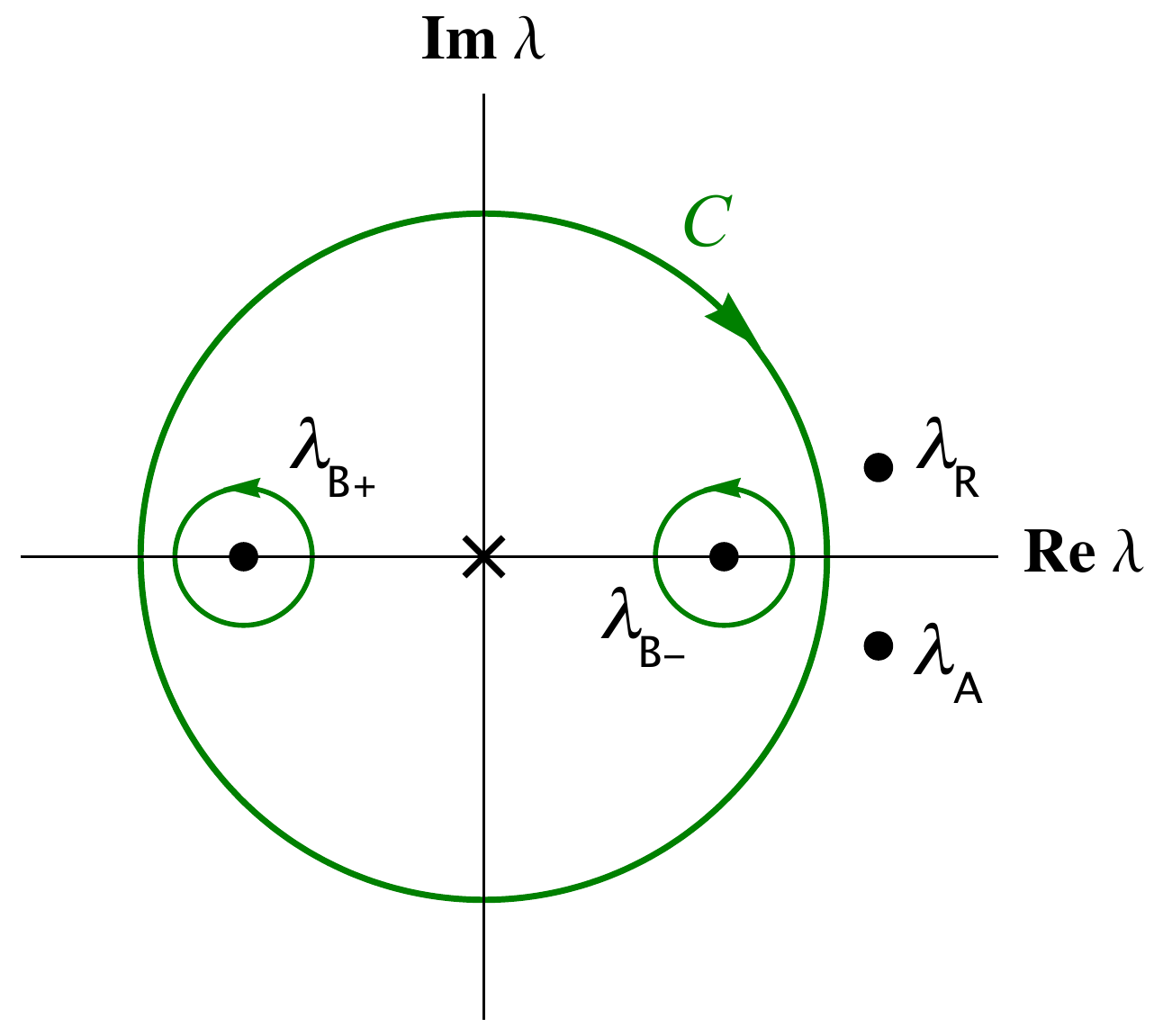}
\hfill
 \includegraphics[width=0.3\textwidth]{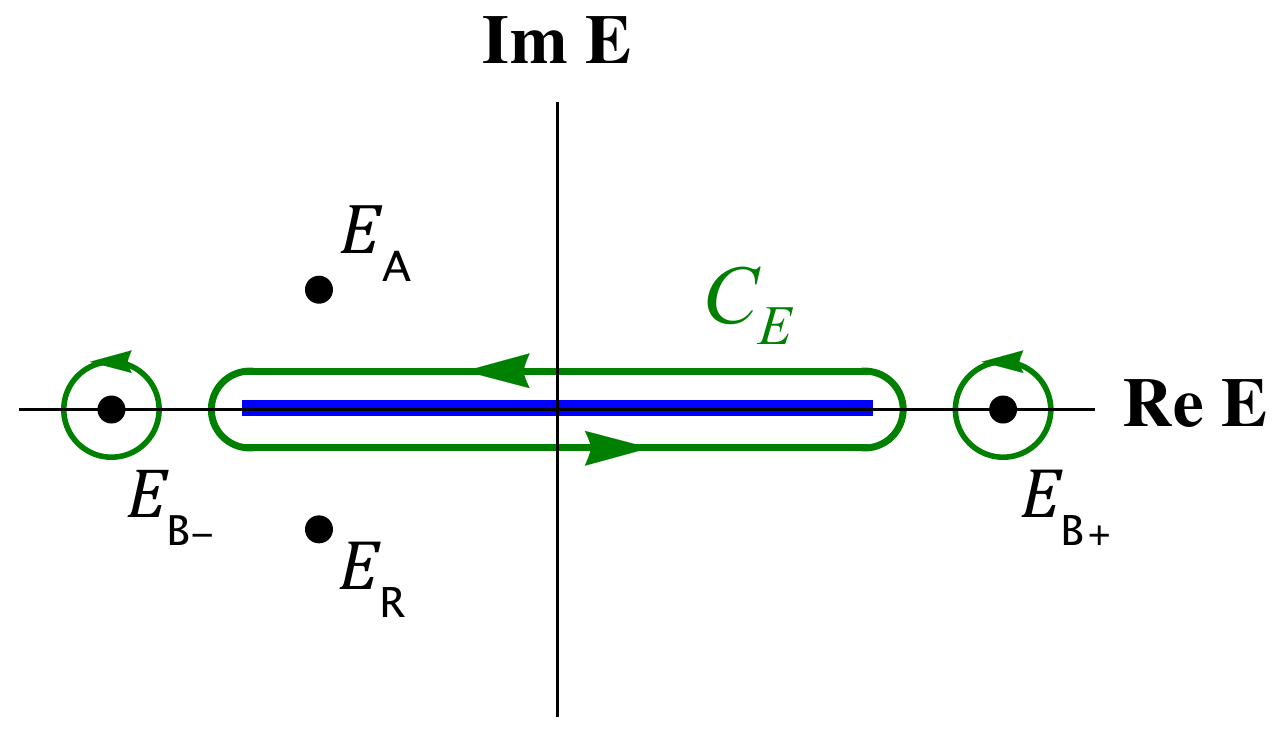}
\hfill
 \includegraphics[width=0.3\textwidth]{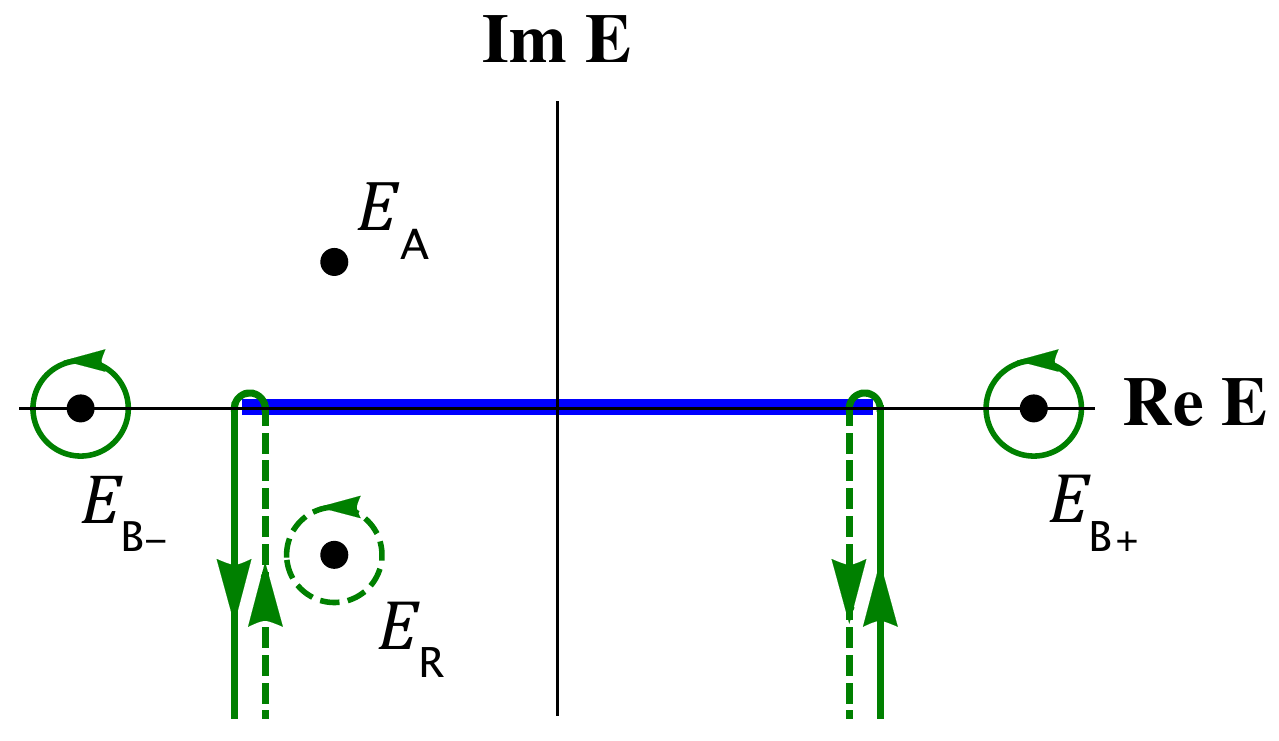}
 \hspace*{0.05\textwidth}
\\
\vspace*{\baselineskip}
\hspace*{0.06\textwidth}(a)\hspace*{0.3\textwidth}(b)\hspace*{0.3\textwidth}(c)\hspace*{0.33\textwidth}
\\
\caption{(a) Integration contour $\mathcal C$ in the complex $\lambda$ plane for the original survival amplitude expression appearing in Eq. (\ref{inf.surv.A}).  (b) Integration contour $\mathcal C_E$ in the complex $E$ plane surrounding the branch cut for $A_{\textrm{bc}} (t)$ in Eq. (\ref{inf.surv.CE}) in addition to the two bound state poles $A^{\textrm{B}\pm}_\textrm{pole} (t)$.  (c)  Deformed integration contour in the complex $E$ plane.  The four contributions extending from $E = \pm 2$ out to infinity in the lower half $E$ plane give rise to the inverse power law term in Eq. (\ref{inf.surv.P.LT}).  The dashed lines represent contour contributions residing in the second Riemann sheet.
}
\label{fig:contour} 
\end{figure*}

\begin{widetext}

We assume that the system is initially in the state in which the dot $| d \ket$ is fully occupied.  The survival probability $P(t) = \left| A (t) \right|^2$ is given as the square modulus of the survival amplitude $A(t) = \bra d | e^{-iHt} | d \ket$, which in the context of the present formalism is written as a sum over the contributions from each eigenstate in the form
\begin{eqnarray}
  A(t) 
     &	= & \bra d | e^{-iHt} | d \ket
     						\nonumber  \\
     &	= & \frac{1}{2 \pi i} \sum_{j = \{ B\pm,A,R \}} 
		\int_\mathcal{C} d \lambda \left( - \lambda + \frac{1}{\lambda} \right) 
			\exp \left[ i \left( \lambda + \frac{1}{\lambda} \right) t \right]
				\bra d | \psi_j \ket \frac{\lambda_j}{\lambda - \lambda_j} \bra \tilde{\psi}_j | d \ket
	.
\label{inf.surv.A}
\end{eqnarray}
The contour $\mathcal C$ is shown in Fig. \ref{fig:contour}(a).
Note that this expression is derived for a generic tight-binding model in Ref. \cite{HO14} and for the present model in Ref. \cite{OH17A}.

\end{widetext}

\begin{figure*}
\hspace*{0.05\textwidth}
 \includegraphics[width=0.3\textwidth]{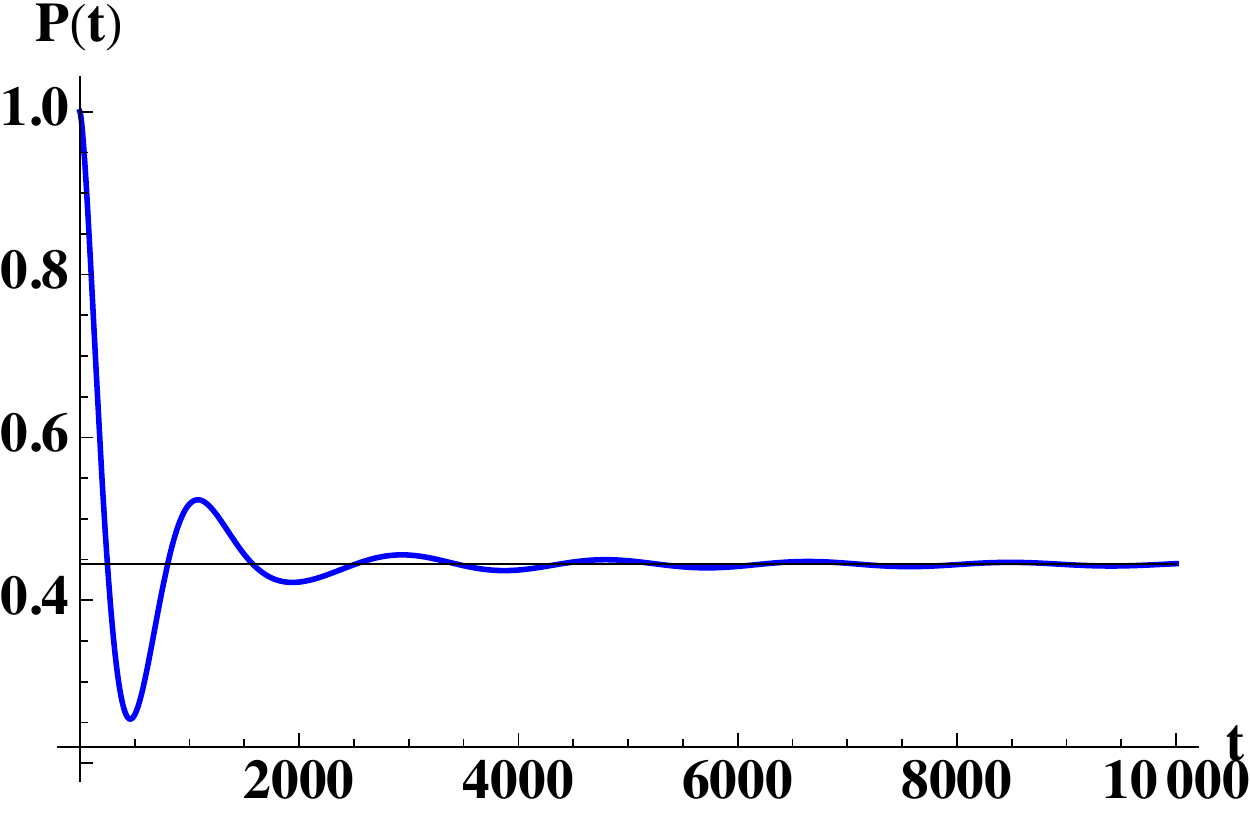}
\hfill
 \includegraphics[width=0.3\textwidth]{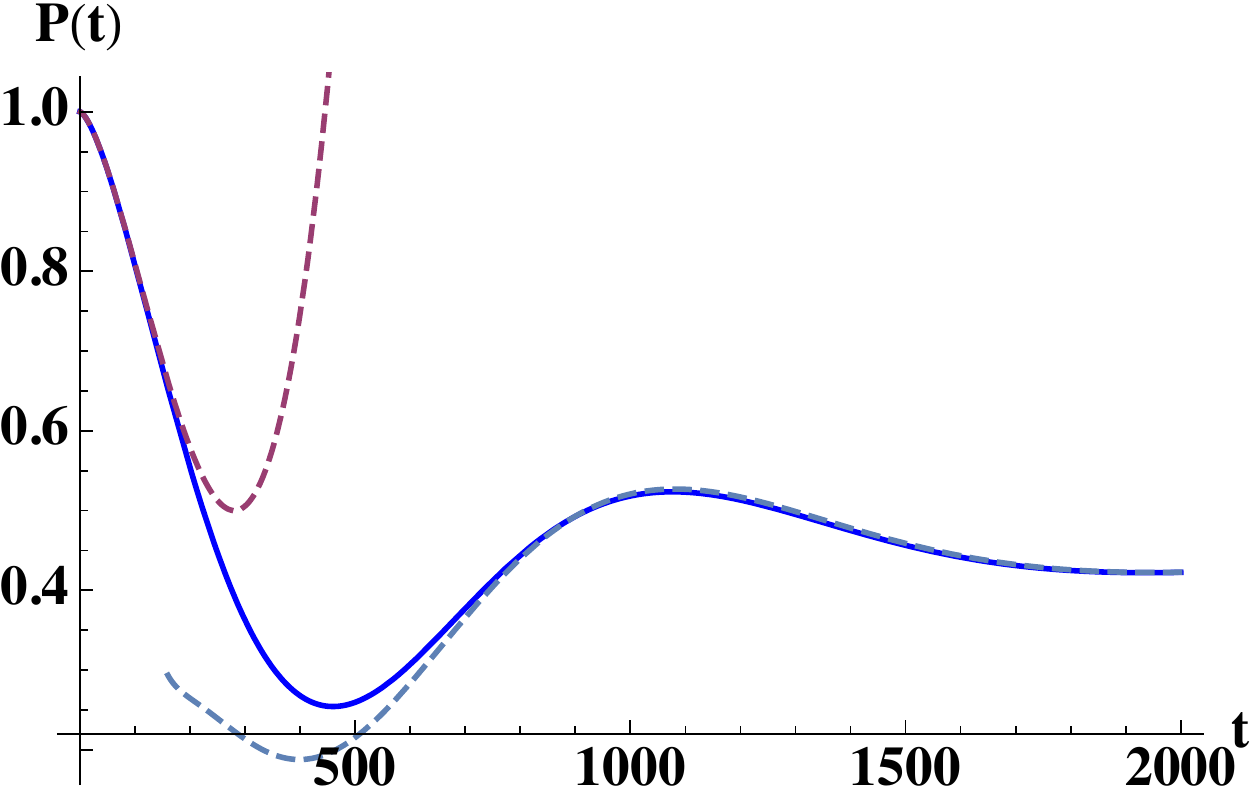}
\hfill
 \includegraphics[width=0.3\textwidth]{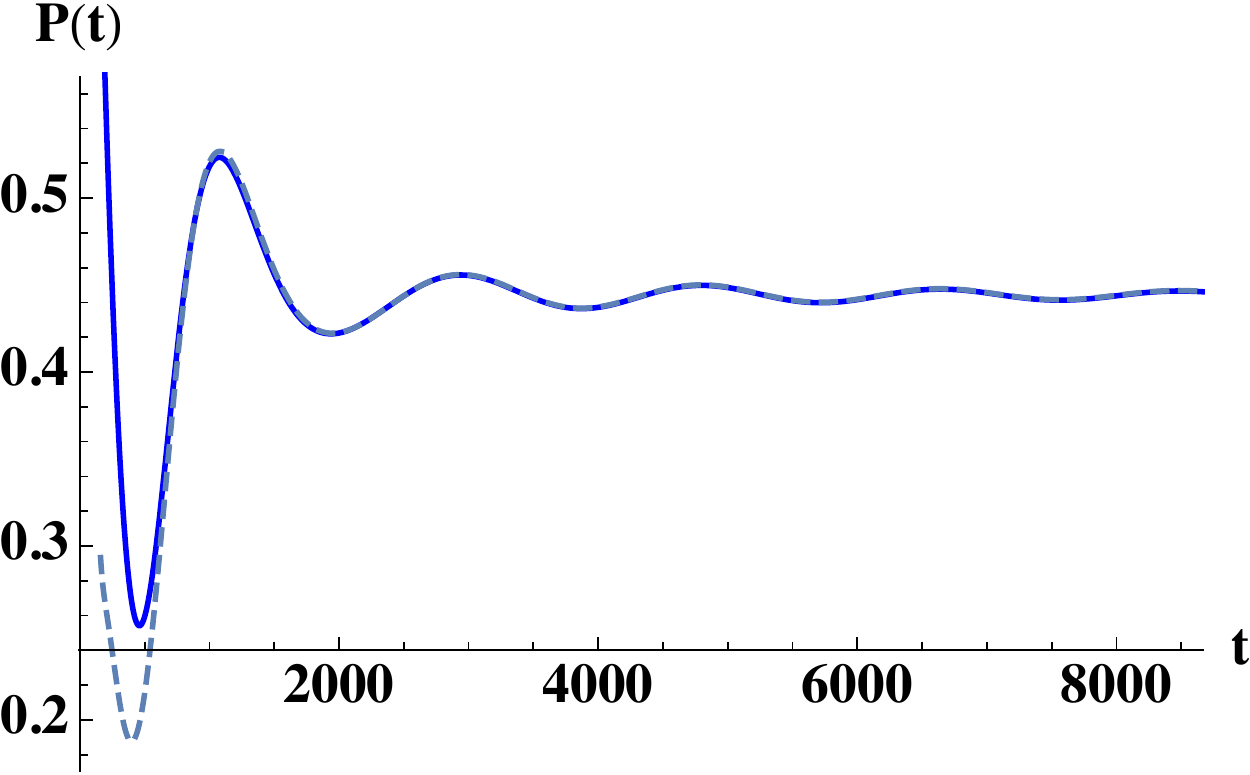}
 \hspace*{0.05\textwidth}
\\
\vspace*{\baselineskip}
\hspace*{0.06\textwidth}(a)\hspace*{0.3\textwidth}(b)\hspace*{0.3\textwidth}(c)\hspace*{0.33\textwidth}
\\
\caption{Survival probability near the anomalous EP for the case $g = 0.02$. (a) The full evolution including the long-time decay behavior.  
(b) Close-up view including the intermediate time zone $1 < t \ll 1/g^{4/3}$ during which the influence of the EP is most strongly pronounced, as shown by the red dashed curve representing the approximation from Eq. (\ref{inf.surv.P.init}).  The transition to the late-time approximation Eq. (\ref{inf.surv.P.LT}) [light blue dotted curve] can also be seen.
(c) Fuller view on the late-time evolution. 
}
\label{fig:EP3.surv} 
\end{figure*}

In Fig. \ref{fig:EP3.surv}(a) we show the evolution for the case $g = 0.02$, very near the threshold EP.   We note that the dynamics can be broken down into several regimes: the earliest (and very brief) quantum-Zeno timescale $t \ll 1/|\epsilon_d|$ yields the usual short-time parabolic decay of the form $P(t) \approx 1 - g^2 t^2$ \cite{SudarshanZeno,SudarshanZeno2,KK96,RaizenZeno1}.  This is quickly followed by an intermediate timescale $1 < t \ll 1/g^{4/3} $, during which most of the decay occurs;
it is during this period that the influence of the anomalous exceptional point on the dynamics is most pronounced, as detailed below.  A dip then appears in the probability that transitions into a decaying oscillatory phase, which in turn gradually settles asymptotically to an incomplete or fractional decay.  The fractional decay is due to population trapping in the bound state $E_{{\textrm B}-}$ near the lower threshold.  Note that the influence of the upper bound state $E_{{\textrm B}+}$ is negligible in this situation.

\begin{widetext}

\subsection{Intermediate-time scale dynamics} \label{sec:time.int}

To evaluate the exceptional point influence on the dynamics on the key timescale, we focus on the contributions to the integral in Eq. (\ref{inf.surv.A}) coming from the three coalescing states $j=\textrm{B}-, \textrm{A, R}$.  Changing the integration variable from $\lambda$ to $E = - \lambda - 1/\lambda$, we can rewrite the outer contour in Fig. \ref{fig:contour}(a) as an integration over the branch cut in the energy plane, as in Fig. \ref{fig:contour}(b).  
Then after neglecting the contribution from the upper bound state $\textrm{B}+$
the total survival amplitude can be written
\beq
  A(t) \approx A_{\textrm{bc}} (t) 
  					+ A^{\textrm{B}-}_\textrm{pole} (t)
\label{inf.surv.total}
\eeq
in which the branch cut contribution now takes the form
\beqa
  A_{\textrm{bc}} (t) & = &
 	- \frac{1}{2 \pi} \sum_{j = \{ \textrm{B}-,\textrm{A,R} \}} \lambda_j \bra d | \psi_j \ket^2
		\int_{\mathcal{C}_E} dE e^{-i E t} \frac{\sqrt{1 - E^2/4}}{E - E_j}
					\label{inf.surv.CE}			\\
 &	= & - \frac{1}{\pi} \sum_{j = \{ \textrm{B}-,\textrm{A,R} \}} \lambda_j \bra d | \psi_j \ket^2
		\int_{-2}^{2} dE e^{-i E t} \frac{\sqrt{1 - E^2/4}}{E - E_j}
	,
\label{inf.surv.E}
\eeqa
and the pole from the lower bound state is evaluated as
\beq
  A^{\textrm{B}-}_\textrm{pole} (t)
  	= \bra d | \psi_{\textrm{B}-} \ket^2 \left(1 - \lambda_{\textrm{B}-}^2 \right) e^{- i E_{\textrm{B}-} t}.
\label{inf.surv.pole}
\eeq
In Eq. (\ref{inf.surv.CE}) the contour $\mathcal C_E$ surrounds the branch cut in the complex $E$ plane as shown in Fig.  \ref{fig:contour}(b).
At this stage, we might consider applying the expansions for the eigenvalues near the EP from Sec. \ref{sec:EP} to evaluate the branch cut contribution $A_{\textrm{bc}} (t)$ (which would be similar to the approach from Ref. \cite{GO17}).  However, doing so, we find evaluating the pole at the band edge to be a challenge.  Instead, we rewrite the pole appearing in $A_{\textrm{bc}} (t)$ as a second integration in the form
\beq
  A_{\textrm{bc}} (t) = 
  	- \frac{i}{\pi} \sum_{j =  \{ \textrm{B}-,\textrm{A,R} \}} \lambda_j \bra d | \psi_j \ket^2
		\int_0^\infty d \tau \int_{-2}^{2} dE e^{-i E t} e^{i (E - E_j) \tau} \sqrt{1 - E^2/4}
	.
\label{inf.surv.E.1}
\eeq
Again transforming the integration variable from the $E$ plane to the $k$ plane according to $E = - 2 \cos k$ allows us to rewrite the inner integration in terms of a Bessel function $J_1 (x)$ as
\beq
  A_{\textrm{bc}} (t) = 
  	- i \sum_{j =  \{ \textrm{B}-,\textrm{A,R} \}} \lambda_j \bra d | \psi_j \ket^2
		\int_0^\infty d \tau e^{-i E_j \tau} \frac{J_1 \left[ 2 \left( t - \tau \right) \right]}{ t - \tau}
	.
\label{inf.surv.k}
\eeq
We note that the information about the branch cut is now encoded in the Bessel function $J_1(2 t)$.
Making a final integral transformation to the variable $t^\prime = t - \tau $ one can show that the resonance and anti-resonance contributions to $A_{\textrm{bc}} (t) $ take the form \cite{OH17A,GO17}
\beq
  A_{\textrm{bc},j} (t) =
		\bra d | \psi_j \ket^2 e^{-i E_j t} \left(
				1 - i \lambda_j \int_0^t dt^\prime e^{i E_j t^\prime} \frac{J_1 (2 t^\prime)}{t^\prime}
				\right)
\label{inf.surv.trans.RA}
\eeq
for $j = \{ R, A \}$, while the bound state contribution instead evaluates as
\beq
  A_{\textrm{bc, B}-} (t) =
		\bra d | \psi_{\textrm{B}-} \ket^2 e^{-i E_{\textrm{B}-} t} \left(
			\lambda_{\textrm{B}-}^2 
			- i \lambda_{\textrm{B}-} 
				\int_0^t dt^\prime e^{i E_{\textrm{B}-} t^\prime} \frac{J_1 (2 t^\prime)}{t^\prime}
				\right)
	.
\label{inf.surv.trans.B}
\eeq
Finally, combining Eqs. (\ref{inf.surv.trans.RA}) and (\ref{inf.surv.trans.B}) with the pole contribution in Eq. (\ref{inf.surv.pole}) we obtain the total survival amplitude [Eq. (\ref{inf.surv.total})] in the form
\beq
  A(t)	\approx  \sum_{j = \{ \textrm{B}-,\textrm{A,R} \}} 
  		\bra d | \psi_j \ket^2 e^{-i E_j t} \left(
				1 - i \lambda_j \int_0^t dt^\prime e^{i E_j t^\prime} \frac{J_1 (2 t^\prime)}{t^\prime}
				\right)
	,
\label{inf.surv.trans}
\eeq
in which the contributions from the three states involved with the EP all appear on an equal footing.  This enables the following analysis.
\end{widetext}

We are now in a position to apply our near-threshold expansions to study the influence of the EP on the evolution during the key timescale.  Introducing the small parameter $\beta \equiv \left( g^2 / 2 \right)^{1/3} $ we can treat $\beta^2 t$ as the smallest non-negligible quantity during the timescale $1 < t \ll 1/\beta^2$ (in which $1/\beta^2 \sim g^{-4/3}$).  For the first term in Eq. (\ref{inf.surv.trans}) we can then expand the exponential and apply the expansions near the EP from Eqs. (\ref{E.B}), (\ref{E.RA}), (\ref{norm.B.d})
and (\ref{norm.RA.d}) to find
\beq
  \sum_{j =  \{ \textrm{B}-,\textrm{A,R} \}} \bra d | \psi_j \ket^2 e^{-i E_j t}
	\approx e^{2it} \left( 1 - \frac{\beta^3 t^2}{2} \right)
	.
\label{inf.surv.trans.1}
\eeq
The eigenvalue expansion has yielded in the second term the factor $t^{2} e^{2it}$, which one would expect in the vicinity of an effective third-order exceptional point.  However, this term will ultimately cancel in the final expression for the survival amplitude.

Next, we perform a similar expansion for the second term in Eq. (\ref{inf.surv.trans}), and then find it useful to rewrite the resulting integrals over the Bessel function according to
\beq
  \mathcal{K}_n (t)
  	= \int_0^t dt^\prime e^{-2 i t^\prime} (t^{\prime})^{n-1} J_1 (2 t^\prime)
	.
\label{K.n}
\eeq
We then use the formulae
\beqa
  \mathcal{K}_0 (t) & = & i \left[ -1 + e^{-2it} \left( J_0 (2t) + i J_1 (2t) \right) \right]
  	\\
  \mathcal{K}_1 (t) & = & \frac{1}{2} \left[ 1 - e^{-2it} \left[ \left(1 + 2it \right) J_0 (2t) - 2t J_1 (2t) \right] \right]
    	\\
  \mathcal{K}_2 (t) & = & \frac{t}{3} e^{-2it} \left[ -it J_0 (2t) + \left( i + t \right) J_1 (2t) \right]
\label{K.n.J}
\eeqa
and finally expand the Bessel functions for $t > 1$ according to
\beqa
  J_0 (2t) & \approx & \sqrt{\frac{1}{\pi t}} \cos \left( 2t - \pi/4 \right)
  	,
	\label{J.n.approx.1}
	\\
  J_1 (2t) & \approx & \sqrt{\frac{1}{\pi t}} \sin \left( 2t - \pi/4 \right)
  	,
\label{J.n.approx.2}
\eeqa
to find the approximation for the second term of Eq. (\ref{inf.surv.trans}) as
\beqa
  - i \sum_{j = \{ \textrm{B}-,\textrm{A,R} \}} \lambda_j \bra d | \psi_j \ket^2 e^{-i E_j t} 
  	 \int_0^t dt^\prime e^{i E_j t^\prime} \frac{J_1 (2 t^\prime)}{t^\prime}
	 			\nonumber	\\
	 	\approx e^{2it} \left( \frac{\beta^3 t^2}{2} 
			- \frac{4 \beta^3 t^{3/2}}{3 \sqrt{\pi}} e^{-i \pi/4} \right)
	.
\label{inf.surv.trans.2}
\eeqa
Notice that the first term of this expression will cancel with the second term of Eq. (\ref{inf.surv.trans.1}), as predicted.  Further, note that the second term here in (\ref{inf.surv.trans.2}) contains the factor $t^{3/2} e^{2it}$, which can be understood as itself consisting of two factors, one from the exceptional point and the other from the threshold.  The former is the factor $t^{2} e^{2it}$ coming from the eigenvalue expansion near the effective third-order EP that we have already encountered.  The second factor $t^{-1/2}$ appeared from expanding the integral over the Bessel function in Eqs. (\ref{J.n.approx.1}) and (\ref{J.n.approx.2}), which is the contribution from the band edge, similar to Refs. \cite{GPSS13,GO17}.

Putting the two terms Eqs. (\ref{inf.surv.trans.1}) and (\ref{inf.surv.trans.2}) together, we obtain our approximation for the time evolution integral Eq. (\ref{inf.surv.trans}) as 
\beq
  A(t) \approx e^{2it} \left( 1 - \frac{2 g^2 t^{3/2}}{3 \sqrt{\pi}} e^{-i \pi/4} \right)
\label{inf.surv.A.init}
\eeq
after again replacing $\beta = \left( g^2 / 2 \right)^{1/3} $ with the physical parameter $g$.  Finally, the resulting approximation for the survival probability itself is given by
\beq
  P(t) \approx 1 - \frac{g^2 t^{3/2}}{3 \sqrt{2 \pi}}
  	.
\label{inf.surv.P.init}
\eeq
This approximation is shown by the dashed curve in Fig. \ref{fig:EP3.surv} (b), which we see captures the dynamics quite well during the period in which the majority of the decay occurs.


\subsection{Long-time scale dynamics} \label{sec:time.long}

Next, we evaluate the dynamics in the latter stages of the evolution.  
As a general rule-of-thumb, we expect that the branch point effect will come to dominate the dynamics as time progresses \cite{Khalfin,RHM06,Fonda,Muga_review,CrespiExpt,GO17,GPSS13,SZMG17}, which often leads to inverse power law decay  as $t \rightarrow \infty$.
To evaluate this, we return to the expression for the survival amplitude near the threshold $A(t) \approx A_{\textrm{bc}} (t) + A_{\textrm{B}-} (t)$ from Eq. (\ref{inf.surv.total}).  We evaluate the branch-cut term $A_{\textrm{bc}} (t)$ by deforming the integral contour $\mathcal C_E$ in Eq. (\ref{inf.surv.CE})
by dragging it into the lower half of the complex energy plane as shown in Fig. \ref{fig:contour}(c).  In doing so we pick up a pole at the resonance eigenvalue, which is evaluated\footnote{Note that this is different from our previous strategy to determine the EP dynamics, in which we instead simplified the integral by effectively removing the pole.} near the threshold as $A^{\textrm{R}}_\textrm{pole} (t) =  \bra d | \psi_{\textrm{R}} \ket^2 \left(1 - \lambda_{\textrm{R}}^2 \right) e^{- i E_{\textrm{R}} t} \approx 2 e^{- i E_{\textrm{R}} t} / 3 + O(g^{2/3})$
after applying the expansions Eqs. (\ref{norm.RA.d}) and (\ref{E.RA}).  Similarly, we can expand for the bound-state pole from Eq. (\ref{inf.surv.pole}) to find $A^{\textrm{B}-}_\textrm{pole} (t) \approx 2 e^{- i E_{\textrm{B}-} t} / 3 + O(g^{2/3})$.  The remaining integration contour in Fig. \ref{fig:contour}(c) extends from the branch points at $E = \pm 2$ out to infinity in the lower half plane; this contribution yields the typical inverse power law decay with $t^{-3/2}$ dependence in the amplitude (see also Refs. \cite{GPSS13,SZMG17}).  Putting these three pieces together the survival probability in this case follows
\beq
  P (t) \approx 
  	 \left| \frac{2}{3} e^{-i E_{\textrm{B}-} t} +  \frac{2}{3} e^{-i E_\textrm{R} t}
	 	- \frac{e^{i \pi/4} e^{2it} }{\sqrt{\pi} g^2 t^{3/2} } \right|^2
	.
\label{inf.surv.P.LT}
\eeq
This is shown as the light blue dotted curve in Fig. \ref{fig:EP3.surv} (c), which focuses on the late-time evolution of the decay.  We note that this approximation picks up around the first minimum, just as the oscillatory dynamics kick in.  The three cross-terms in Eq. (\ref{inf.surv.P.LT}) each give a decaying oscillation with frequency $\sim g^{4/3}$.  However, the resonance contribution with lifetime $\Gamma^{-1} \sim g^{-4/3}$ [about $ \sim 184.2$ for $g=0.02$ as in Fig. \ref{fig:EP3.surv} (c)] has mostly died out after the first oscillation, so that the dominant contribution comes from the bound-state/power-law cross term, with oscillation frequency $E_{\textrm{B}-} + 2 = g^{4/3}/2^{2/3}$.  Finally, in the long-time limit, the dot survival probability settles down to the asymptotic occupation probability for the bound state, given simply by $P_{\infty} = \lim_{t \rightarrow \infty} | A^{\textrm{B}-}_\textrm{pole} (t) |^2 = 4/9$.  This is shown by the horizontal black line in Fig. \ref{fig:EP3.surv} (a).


\section{Discussion} \label{sec:conc}

In this work, we have shown the existence of an anomalous-order exceptional point at threshold (band edge) in a simple 1-D continuum model with an attached quantum emitter and examined its influence on the decay dynamics when the emitter is tuned to the threshold energy.  Although we have shown this in the specific model in Eq. (\ref{inf.ham}), we claim that it is a much more ubiquitous effect.  Let us consider a generic model for a simple open quantum system of the form
\begin{equation}
H =
\epsilon_q q^\dag q
 + \int dk \, E_k \, c_k^\dag c_{k}
 + g \int dk \, ( v_k \, c_k^\dag q +  v_k^* \, q^\dag c_k)  
    ,
\label{gen.ham}
\end{equation}
consisting of the quantum emitter $q^\dag$ coupled to the continuum mode $c_k^\dag$.  Note that our original Hamiltonian Eq. (\ref{inf.ham}) can be placed in this form after applying a simple Fourier transform.  Here the continuum $E_k$ ranges from a lower threshold $E_{\textrm{th}-}$ to an upper threshold $E_{\textrm{th}+}$ (the latter of which might appear at infinity).  We claim that there are two conditions to be placed on this model in order for the anomalous threshold EP to be realized.  
These conditions are most naturally expressed after writing the self-energy function $\Sigma(z)$ as
\begin{equation}
  \Sigma (z) 
	= g^2  \int d E \rho(E) \frac{ \left| v(E) \right|^2 }{z - E}
	,
\label{gen.self.omega}
\end{equation}
in which $\rho(E) = \partial k / \partial E$ is the density of states (DOS) function 
and we have re-written $v(E) = v_{k(E)}$ by inverting $E_k$. 
(We note the appearance of the so-called {\it reservoir structure function} $\rho(E)  \left| v (E) \right|^2$ in this expression, as previously defined in the literature \cite{DG03,LonghiPRA06}.) 
The first of the two conditions 
is that the density of states must contain the square-root divergence
\beq
  \rho (E) \sim \frac{1}{\sqrt{E_\textrm{th}-E}}
\eeq
at either threshold $E_\textrm{th}$, as occurred for our specific model in Eq. (\ref{inf.DOS}).
This 
is indeed the standard form for the van Hove singularity in 1-D systems \cite{vanHove,Mahan,Economou}.  Then the second condition appears on $v (E)$, which simply requires that this quantity should be non-singular at the threshold $E = E_\textrm{th}$.  Assuming these two conditions hold, then as shown in App \ref{app:gen}
the self-energy function will reproduce the square-root divergence, as occurred for our specific model in Eq. (\ref{inf.self}), and hence the spectrum will contain the triple-level convergence illustrated in Fig. \ref{fig:plane.EP} with energy eigenvalues that yield the characteristic $g^{4/3}$ Puiseux expansion, similar to Eqs. (\ref{E.B}) and (\ref{E.RA}) in the main text [in the case of the general model, the Puiseux expansion follows directly from Eq. (\ref{app.gen.g43})].  Physically, the quantity $g^{4/3}$ appears in the inverse period of the oscillations in Fig. \ref{fig:EP3.surv}.
We mention that some quantities, such as the norm in Eqs. (\ref{norm.B.d}) and (\ref{norm.RA.d}), exhibit a slightly more general expansion in terms of $g^{2/3}$.

Illustrating this universality, we note that a similar picture for the dynamics shown in Fig. \ref{fig:EP3.surv}
has previously appeared in the literature \cite{KKS94,John94,LNNB00} in the context of spontaneous emission from atoms near a band edge within photonic band gap (PBG) materials. In Ref. \cite{LNNB00} the authors note ``the peculiar power of $\frac{2}{3}$ being one of the signatures of the unconventional nature of the PBG environment.''   In the present work, we have revealed that the exceptional point is the underlying mechanism that shapes the dynamics near such a band edge, and that its Puiseux expansion is the origin of the ``peculiar power of $\frac{2}{3}$.''  (See Refs. \cite{PTG05,TGP06,CCCR16,SWGTC16,Joe18} for other models in which a similar expansion appears.)  Indeed, in Refs. \cite{John90,John94} the authors highlight the existence of a ``doublet'' consisting of an atom-photon bound state and a resonance when an atom is located near the photonic band edge; however, the presence of an anti-resonance is not noted.  We have shown that it is the eigenvalue convergence and (partial) coalescence involving all three eigenstates that underlies the physics in this situation.

While the anomalous-order exceptional point should be a fairly universal feature, it should be considered that in this work we have analyzed the problem only at the level of the Hamiltonian dynamics, which cannot account for processes like quantum jumps.  Several quite recent works have revealed that the parametric location and dynamical characteristics of the exceptional points can become modified 
when treating the problem at the level of the Liouvillian formalism with Lindblad terms that can account for such processes \cite{HatanoEP3,BreuPet,LEP,Joglekar_periodic,KBH21}.  However, recent experiments 
in circuit quantum electrodynamics have to some extent circumvented this issue through the process of data post-selection, in which trials that result in quantum jumps are eliminated from the final analysis \cite{Murch19,Murch21}.  
We propose that experimental observation of the features of the anomalous exceptional point might be achieved by a modified version of these experiments in which a superconducting qubit is embedded in a waveguide (instead of a cavity as in the original experiments \cite{Murch19,Murch21}) with the qubit transition frequency tuned to the lowest waveguide cutoff mode.


We close the paper with the following two comments aimed at placing our results in the context of the wider literature.  


\subsection{Comment on the anomalous order of the EP} \label{sec:conc.order}

We emphasize that one peculiar mathematical aspect of the present results is the sharp distinction in the near-threshold spectral properties of the system in the case of small coupling as opposed to the case in which the coupling actually vanishes.  In the former case, the system behaves precisely as if there were an EP3 at the threshold, whereas in the latter case there instead appears an exact EP2.  To connect these seemingly incongruent pictures, we showed in App. \ref{app:JB} that a reorganization of the eigenstates occurs in the limit in which the coupling vanishes such that two linear combinations of the three relevant eigenstates converge on the state of the decoupled quantum emitter, while a third linear combination instead merges with the continuum at the threshold.  We mention that this resolution of the problem was partly inspired by Ref. \cite{Hashimoto}, in which Hashimoto, {\it et al}, introduced a novel basis near an exceptional point such that the system eigenvectors can connect continuously with the Jordan block representation directly at the EP, which we referred to in our internal discussions as {\it Hashimoto's representation}.  

It is interesting to note that the mismatch between the Puiseux expansion and the exact order of the EP in the present case is rather different than what has appeared previously in the literature.  First, in the most typical case, in the vicinity of an order $N$ exceptional point the relevant eigenvalues can be expanded in a Puiseux series written in terms of an $N$th order root.  However, in Ref. \cite{GraefeEP3} the authors point out the existence of a special case in which the eigenvalues are instead organized into distinct subgroups (called {\it cycles}), which have their own separate Puiseux expansions that are lower order than $N$.  For example, they demonstrate that an EP3 can occur for which two coalescing eigenvalues can be expanded in a square root near the EP3 (order 2 Puiseux expansion), while the other coalescing eigenstate instead takes the form of a Taylor series (which can be viewed as an order 1 Puiseux expansion).  Notice then that the order of the two cycles adds up to the order of the EP itself $3 = 2 + 1$.  We emphasize that here the order of the individual Puiseux expansions is always {\it lower} than the order of the EP. 

In contrast, in our case not only is there a mismatch between the order of the EP and that of the root appearing in the Puiseux expansion, but the Puiseux expansion is instead {\it higher} order than the EP itself.  Further, this mismatch is resolved in a fundamentally different way: instead of subgroups of eigenvalues with their own Puiseux expansions, the eigenstates in our case form subgroups with different coalescing behaviors (two linear combinations of eigenstates coalesce on the discrete quantum emitter while a third instead merges with the continuum). 

The resolution to the mismatch in the present case points to the reason that a more general type of exceptional point is allowed in our model in the first place.  The key point is that in Ref. \cite{GraefeEP3} the authors only consider {\it finite}, non-Hermitian matrices, in contrast to our model, which incorporates an eigenvalue continuum.  Hence we conclude that the presence of continuous eigenvalue spectra in a given Hamiltonian permits a wider diversity of exceptional point types. 


\subsection{Wider context of near-threshold non-Markovian dynamics in 1-D systems} \label{sec:conc.dynamics}

While here we have analyzed a situation in which the eigenvalues of three states converge directly on the continuum threshold, we note that the dynamical influence of individual states appearing near the threshold has previously been studied in the literature in 1-D systems.  In Refs. \cite{Jittoh,GCV06} it has been shown that a resonance with a near-threshold eigenvalue can lead to full-time non-exponential decay.  Meanwhile, in Refs.\cite{LonghiPRA06,DBP08,GPSS13,GNOS19} it is shown that a virtual bound state (or anti-bound state) merging directly with the continuum threshold can also result in full-time inverse power law decay.  More generally, the virtual state appearing in the vicinity of the threshold introduces a timescale that characterizes the power law decay  \cite{GPSS13,GNOS19}.  This timescale is inversely proportional to the energy gap between the virtual state eigenvalue and the threshold, and it divides the dynamics into an intermediate-time zone during which the survival amplitude (survival probability) follows $1/t^{1/2}$ ($1/t$) decay, and a long-time zone in which the amplitude (probability) instead falls off as $1/t^{3/2}$ ($1/t^3$) \cite{GPSS13,GNOS19}.  (We emphasize that a localized bound state near the threshold introduces the same timescale as the virtual bound state, but in this case the decay dynamics are to some extent obscured by the incomplete decay resulting from the bound state itself \cite{GPSS13}.)

Meanwhile, the dynamics associated with multiple ($N$) eigenstates that coalesce at a real eigenvalue that itself appears near the threshold is fundamentally different from, yet also related to the above cases.  In Ref. \cite{GO17} the case of two virtual states coalescing before forming a resonance/anti-resonance pair is considered.  In that paper, when the eigenvalue at which the two states coalesce is near the threshold the picture is somewhat similar to that of the lone near-threshold virtual state, except that the the intermediate timescale dynamics is replaced with a decay of the form $1 - C t^{1/2} \ $ \cite{GO17}.  Meanwhile, in the present work, we have shown that a triplet of resonance, anti-resonance and bound eigenstates converging directly at the threshold results in decay on the intermediate timescale of the form $1 - C t^{3/2}$.  In either case, the intermediate dynamics are replaced by the typical $1/t^3$ decay on the asymptotic timescale, which matches with the single virtual state case.  Attempting to generalize from this picture, we propose that the intermediate timescale dynamics for $N$ converging eigenstates with real eigenvalue near the threshold might be expected to contain a term with the time dependence $ \sim t^{N-3/2}$.  This can be understood as appearing after taking the square modulus of the amplitude containing the terms
$(1 - \textrm{const} \frac{1}{t^{1/2}} \times t^{N-1} )e^{-i \bar{E}_\textrm{EP} t} $, in which the factor 
$t^{N-1} e^{-i \bar{E}_\textrm{EP} t}$ in the second term arises from an $N$th-order pole appearing due to the $N$ converging states, and the factor $1/t^{1/2}$ 
results from the influence of the continuum threshold, just as in the case of the single near-threshold virtual state discussed in the previous paragraph.  However, for the anomalous-order EP in the present case, the number $N$ seems to be equal to the number of converging levels ($N=3$) rather than the technical order of the EP, which suggests that one might have to consider the situation in which the EP occurs directly at the threshold on a case-by-case basis.

The results for the near-threshold dynamics in 2-D and 3-D systems reported in Refs. \cite{GTC2D1,GTC2D2,GTC3D} suggest that one probably must consider these cases separately, as well. 
Another interesting case that one could consider would be that of multiple eigenstates converging on a {\it complex} eigenvalue near a threshold, which would also likely be modified somehow from the above scenario.


\begin{acknowledgements} 
We thank Eva-Maria Graefe, Kazunari Hashimoto, Kater Murch and Dvira Segal for enlightening discussions related to this work.  
S. G. acknowledges support from the Japan Society for the Promotion of Science (JSPS) under KAKENHI Grant No. JP18K03466 and from the Research Foundation for Opto-Science and Technology. 
N. H. acknowledges support from JSPS under KAKENHI Grant No. JP19H00658.
\end{acknowledgements}

\appendix


\section{Eigenstates in the $g \to 0$ limit and connection to the $g \neq 0$ case}\label{app:JB}

In this appendix, we continue our development from Sec. \ref{sec:JB}, illustrating explicitly how to connect the $2 \times 2$ Jordan block appearing in the $g \to 0$ limit with the EP3-like behavior for $g \neq 0$.

First we report the explicit from of the so-called rotation matrix $R$ appearing in Eq. \ref{JB} as 
$R = \{ | \Psi_+ \ket, | \Psi_d \ket, | \Phi_d \ket, | \Psi_- \ket \}$, in which
\beq
   | \Psi_+ \ket = \left( \begin{array}{c}   -1	\\ 0	\\ 1	\\ 0 \end{array} \right)
   ,
\label{Psi.plus}
\eeq
and
\beq
   | \Psi_d \ket = \left( \begin{array}{c}   0	\\ 1	\\ 0	\\ 1 \end{array} \right)
   ,
   | \Phi_d \ket = \left( \begin{array}{c}   0	\\ -1	\\ 0	\\ 0 \end{array} \right)
   ,
   | \Psi_- \ket = \left( \begin{array}{c}   1	\\ 0	\\ 1	\\ 0 \end{array} \right)
   ,
\label{Psi.Phi}
\eeq
are the generalized eigenstates at the threshold EP.
Here  $| \Psi_+ \ket$ in the first line represents the $g \rightarrow 0$ limit for the lone bound state that previously appeared above the upper band edge in the case $g \neq 0$.  For $g \rightarrow 0$, this state takes the eigenvalue $\lambda_+ = -1$.  Comparing the explicit eigenstate $| \Psi_+ \ket = \{ -1, 0, 1, 0 \}^\textrm{T}$ with the generic form Eq. (\ref{inf.Psi}), we see that this state is completely decoupled from the dot as 
$\bra d | \psi_+ \ket = 0$.  This indicates that $| \Psi_+ \ket$ has merged with the continuous spectrum in the $g \rightarrow 0$ limit. 

The three remaining states $| \Psi_d \ket, | \Phi_d \ket$ and $ | \Psi_- \ket$  are those that each share the eigenvalue $\lambda_j = 1$.  This includes $| \Psi_d \ket = \{ 0, 1, 0, 1 \}^\textrm{T}$, which is the uncoupled dot state, and $| \Phi_d \ket = \{ 0, -1, 0, 0 \}^\textrm{T}$, which is the associated pseudo-eigenstate.  Together, these two states satisfy the Jordan-chain relations
\beqa
  \left( A^{-1} B \right) | \Psi_d \ket & = &  | \Psi_d \ket
\label{Psi.d.eqn}	\\
  \left( A^{-1} B \right) | \Phi_d \ket & = &  | \Phi_d \ket + | \Psi_d \ket
  	.
\label{Phi.d.eqn}
\eeqa
Finally, $ | \Psi_- \ket = \{ 1, 0, 1, 0 \}^\textrm{T}$ again represents a state that is completely decoupled from the dot and has joined the continuous spectrum.  While $| \Psi_+ \ket$ merged with the continuum at the upper band edge with eigenvalue $\lambda_+ = -1$ ($E=2$) , $| \Psi_- \ket$ is instead merged at the lower band edge with eigenvalue $\lambda_- = 1$ ($E=-2$).

However, the observant reader may take some discomfort at this point, noting that it is not immediately clear how the three states $| \Psi_\textrm{R} \ket, | \Psi_\textrm{A} \ket$ and $| \Psi_{\textrm{B}-} \ket$ involved in the convergence on $\lambda = 1$ for small $g$ from Sec. \ref{sec:EP} connect with the three (generalized) states $| \Psi_d \ket, | \Phi_d \ket$ and $ | \Psi_- \ket$ appearing in the actual $g \rightarrow 0$ limit.  In particular, the state $ | \Psi_- \ket$ cannot be directly connected with the limiting behavior of the lower bound state $| \Psi_{\textrm{B}-} \ket$ in the same way that $ | \Psi_+ \ket$ simply emerged as the limit of the upper bound state.  Indeed, Eq. (\ref{norm.B.d}) seems to suggest that 
$| \Psi_{\textrm{B}-} \ket$ should just participate in the eigenvalue coalescence with the other two states.

The resolution to this issue comes in the realization that the states $| \Psi_d \ket, | \Phi_d \ket$ and $ | \Psi_- \ket$ can only be obtained as the limit of specific linear combinations of the bound state, resonance state and anti-resonance state.  This approach borrows conceptually from Hashimoto's representation (introduced in Ref. \cite{Hashimoto}), in which an extended Jordan block representation is written such that the eigenvectors away from an exceptional point can connect continuously with those appearing directly at the EP (see also Sec. V of Ref. \cite{KGTP17}).  

As our initial step, we note that in the case $\epsilon_d = -2$ we can use Eq. (\ref{inf.quad.2}) to write 
$g  \lambda_j \bra 0 | \psi_j \ket = (1 - \lambda_j)^2 \bra d | \psi_j \ket $.  Combining this with Eqs. (\ref{norm.B.d}) and (\ref{norm.RA.d}), we can rewrite (\ref{inf.Psi}) as
\beq
  | \Psi_\textrm{j}  \ket
  	= \bra d | \psi_j \ket	
		\left( \begin{array}{c}
			e^{\frac{-2 \pi i \alpha_j}{3}} \left( \frac{g}{4} \right)^{1/3}	 + O(g)	\\
			1											\\
			e^{\frac{-2 \pi i \alpha_j}{3}} \left( \frac{g}{4} \right)^{1/3}	 + O(g)	\\
			1 - e^{\frac{2 \pi i \alpha_j}{3}} \left( \frac{g^2}{2} \right)^{1/3} + O(g^{4/3})  \\ 
	\end{array}
	\right)
	.
\label{inf.Psi.exp}
\eeq
in which $\alpha_j = 0$ for the bound state (with $j = \textrm{B}-$), $\alpha_j = -1$ for the resonance ($j=\textrm{R}$) and $\alpha_j = 1$ for the anti-resonance  ($j=\textrm{A}$).
Using this expression, we first obtain the dot state from the most straightforward linear combination
\beqa
  | \Psi_d \ket
     &	= & \lim_{g \rightarrow 0} 
		\sum_{j=\textrm{B-,R,A}} \frac{1}{3 \bra \textrm{d} | \psi_j \ket} | \Psi_j \ket
							\nonumber	\\
     &	= & \lim_{g \rightarrow 0} \left( \begin{array}{c}  
						O(g)    \\ 1	\\ O(g) 	\\ 1 + O(g)
			\end{array} \right)
	=  \left( \begin{array}{c}  
						0    \\ 1	\\ 0	\\ 1
			\end{array} \right)
	.
\label{Psi.d}
\eeqa
To obtain the partner pseudo-dot state, however, we need to be a little more clever.  From the following linear combination, we can engineer the cancellation of all the leading-order entries in Eq. (\ref{inf.Psi.exp}), so that we get
\beqa
  | \Phi_d^\prime \ket
     &	= & \lim_{g \rightarrow 0} 
		\sum_{j=\textrm{B-,R,A}} 
			\frac{e^{\frac{-2 \pi i \alpha_j}{3}}}{3 \left(g^2/2\right)^{1/3} \bra \textrm{d} 
													| \psi_j \ket} | \Psi_j \ket
							\nonumber	\\
     &	= & \lim_{g \rightarrow 0} \left( \begin{array}{c}  
						O(g^{1/3})    \\ 0	\\ O(g^{1/3}) 	\\ -1 + O(g^{2/3})
			\end{array} \right)
	=  \left( \begin{array}{c}  
						0    \\ 0	\\ 0	\\ -1
			\end{array} \right)
	.
\label{Phi.d}
\eeqa
Note that this pseudo-state is actually different from $| \Phi_d \ket$ appearing in Eq. \ref{Psi.Phi}.  In fact, this pseudo-state satisfies the Jordan-chain relation
\beq
  \left( A^{-1} B \right) | \Phi_d^\prime \ket =  | \Phi_d^\prime \ket - | \Psi_d \ket
  	,
\label{Phi.dd.eqn}
\eeq
which is different from Eq. (\ref{Phi.d.eqn}).  However, a new, equally valid pseudo-state can always be obtained from a previously known one by combining it with the corresponding eigenvector.  Hence, we can easily obtain the original pseudo-state $| \Phi_d \ket$ by writing $| \Phi_d \ket = - | \Phi_d^\prime \ket -  | \Psi_d \ket$.

Finally, we can obtain the state appearing at the lower band edge from the combination
\beqa
  | \Psi_- \ket
     &	= & \lim_{g \rightarrow 0} 
		\sum_{j=\textrm{B-,R,A}} 
			\frac{e^{\frac{2 \pi i \alpha_j}{3}}}{3 \left( g/4 \right)^{1/3} \bra \textrm{d} | \psi_j \ket} | \Psi_j \ket
							\nonumber	\\
     &	= & \lim_{g \rightarrow 0} \left( \begin{array}{c}  
						1 + O(g^{2/3})    \\ 0	\\ 1 + O(g^{2/3}) 	\\ O(g)
			\end{array} \right)
	=  \left( \begin{array}{c}  
						1    \\ 0	\\ 1	\\ 0
			\end{array} \right)
	.
\label{Psi.m}
\eeqa


\section{Triple-level convergence appearing in the generic model}\label{app:gen}

In Sec. \ref{sec:conc} we claimed for the generic model given in Eq. (\ref{gen.ham}) that two conditions are required for the triple-level convergence in Fig. \ref{fig:plane.EP} and anomalous exceptional point from the main text to occur.  One of these conditions was that the continuum must be 1-D, which yields the characteristic van Hove singularity 
$( E_\textrm{th}-E )^{-1/2}$.  The second condition was that the potential $v(E)$ must be non-singular at the threshold $v (E_\textrm{th})$.  We motivate these claims in what follows.


We briefly note that we have applied the rotating-wave approximation in writing our generic model in Eq. (\ref{gen.ham}).  However, Ref. \cite{PTG05} provides an example of a slightly more general model in which counter-rotating terms are retained and yet the threshold EP still occurs.

\subsection{Form of self-energy for the generic model}\label{app:gen.self}

Here we quickly outline the conditions such that the self-energy function from Eq. (\ref{gen.self.omega}) for the generic model reproduces the square-root divergence in the denominator from the DOS, which in turn yields the anomalous EP.
We start by rewriting the self-energy function in terms of an integration in the complex wave vector plane as
\beq
  \Sigma (E) = g^2  \int_{- \infty}^{\infty} dk \frac{ \left| v_k \right|^2 }{E - E_k}
  	.
\label{app.self}
\eeq
We assume that $\Sigma (E)$ is defined first for real $E$ appearing below the continuum threshold 
$E < E_\textrm{th}$ (i.e. a bound state).
Afterwards the result can be analytically continued into the complex domain.  Next we assume a generic dispersion relation of the form $E_k \sim k^2 + E_\textrm{th}$, which might be exact or merely an approximation near the threshold $E_\textrm{th}$.  Then we rewrite $\Sigma(E)$ in the form
\beq
  \Sigma (E) = - g^2  \int_{- \infty}^{\infty} dk \frac{ \left| v_k \right|^2 }{\left( k - k_+ \right) \left( k - k_- \right)}
\label{app.self.kpm}
\eeq
in which $k_\pm = \pm i \sqrt{E_\textrm{th} - E}$.
Since the integrand of this expression vanishes like $1/k^2$ as $k \to +i \infty$, we can close the integration contour in the upper half of the $k$-plane, so that taking the residue at $k_+$ gives
\beq
  \Sigma (E) 
  	= - \frac{ \pi g^2  \left| v_{k_+(E)} \right|^2 }{\sqrt{E_\textrm{th} - E}}
	.
\label{app.self.final}
\eeq
Thus the self-energy function obtained by integration over the 1-D continuum has recreated the square-root divergence from the DOS, which verifies our first condition.

For the second condition, notice that if $v_{k_+}$ contained some singularity at $k_+ = 0$ ($E = E_\textrm{th}$), this would have the effect of disturbing the square root in the denominator and thus the anomalous exceptional point would no longer appear (as occurs for the models in Refs. \cite{DBP08,BCP10,GNOS19}, for example).

Note that a further term that is analytic at $E_\textrm{th}$ might arise depending on the precise form of $v_k$ or integration over a second continuum with a different threshold, for example.  Hence, we can write the self-energy in Eq. (\ref{app.self.final}) in a slightly more general form as
\beq
  \Sigma (E) 
  	= g^2 \Delta (E) + g^2 \frac{\lambda(E)}{\sqrt{E_\textrm{th} - E}}
\label{app.self.gen}
\eeq
in which $\Delta (E)$ and $\lambda(E)$ are both analytic in $E$.  (Notice that this is now more general than the model from the main text.)
One could reasonably suppose that the presence of the $\Delta(E)$ term here might disrupt the occurence of the  triple-level convergence associated with the exceptional point for this more general case.  We show in App. \ref{app:gen.deg} that indeed the triple-level convergence still occurs.


\subsection{Generality of the triple degeneracy at threshold}\label{app:gen.deg}

Here we demonstrate that the form of the self-energy reported in Eq. (\ref{app.self.gen}) indeed yields the triple degeneracy at threshold in the limit of vanishing coupling, as described in the main text and in Fig. \ref{fig:plane.EP}.

The point spectrum for the generic model is obtained from the dispersion equation $E - \epsilon_q  =  \Sigma (E)$, which corresponds to Eq. (\ref{inf.self}) for the model in the main text.  To study the point spectrum in the vicinity of the threshold we specify $\epsilon_q = E_\textrm{th}$ in this equation and use the explicit form of Eq. (\ref{app.self.gen}) to write
\beq
  E - E_\textrm{th} =  g^2 \Delta (E) + g^2 \frac{\lambda(E)}{\sqrt{E_\textrm{th} - E}}
  	.
\label{gen.disp.th}
\eeq
Multiplying through by $x \equiv \sqrt{E_\textrm{th} - E}$ we obtain a cubic equation in $x$ of the form
  $x^3 + g^2 \Delta (E) x + g^2 \lambda (E) = 0$,
which is solved by
\beq
  \sqrt{E_\textrm{th} - E} = \xi_+^{1/3} + \xi_-^{1/3}
\eeq
with
\beq
  \xi_{\pm} = \frac{-g^2 \lambda \pm g^2 \lambda \sqrt{1 + \frac{4 g^2 \Delta^3}{27 \lambda^2}}}{2}
  	. 
\eeq
For small $g \ll 1$, we find that $\xi_- \approx - g^2 \lambda$ is of lower order than 
$\xi_+ \approx g^4 \Delta^3 / 27 \lambda$, which results in
\beq
  \sqrt{E_\textrm{th} - E} \approx - \left( g^2 \lambda \right)^{1/3}
  	.
\label{app.gen.g43}
\eeq
Squaring, then cubing, we immediately see the familiar form $(E - E_\textrm{th})^3 = - g^4 \lambda^2$ from the main text in Sec. \ref{sec:inf.Heff}, which yields the expected triple-level convergence as $g \rightarrow 0$ as well as the Puiseux expansion in terms of $g^{4/3}$.  (See also Refs. \cite{PTG05,LNNB00,KKS94,TGP06,GNHP09,CCCR16,SWGTC16,Joe18} for specific models in which a similar expansion has appeared.)

\end{document}